# Power Minimization in Vehicular Cloud Architecture

Fatemah S. Behbehani, Taisir Elgorashi, and Jaafar M. H. Elmirghani, *Fellow IEEE*

*School of Electronic and Electrical Engineering, University of Leeds, Leeds, LS2 9JT, United Kingdom*

*Abstract*— Modern vehicles equipped with on-board units (OBU) are playing an essential role in the smart city revolution. The vehicular processing resources, however, are not used to their fullest potential. The concept of vehicular clouds is proposed to exploit the underutilized vehicular resources to supplement cloud computing services to relieve the burden on cloud data centers and improve quality of service. In this paper we introduce a vehicular cloud architecture supported by fixed edge computing nodes and the central cloud. A mixed integer linear programming (MLP) model is developed to optimize the allocation of the computing demands in the distributed architecture while minimizing power consumption. The results show power savings as high as 84% over processing in the conventional cloud. A heuristic with performance approaching that of the MILP model is developed to allocate computing demands in real time.

*Index Terms*— **vehicular clouds, edge nodes, power optimization, distributed processing.**

## I. INTRODUCTION

Cloud computing has introduced new possibilities for data processing and storage. The paradigm provides remote services that relieve the end users from handling data processing and storage in their own devices [1]. It reduces the cost for users by eliminating the need to deploy and maintain hardware and software resources. On the providers side, the costs of provisioning cloud services are expected to be well compensated through the profit of the growing cloud services. The demand for cloud services is growing exponentially with 30-40% annual traffic growth [2]. Cost and power consumption of cloud computing and data centers are significantly contributing to the total cost and power consumption in the Information and Communication Technology (ICT) field [1-4].

In addition, large data centres tend to be located away from the end users, which increases the latency and power consumption of the networks interconnecting users to the cloud. Also, some applications such as smart city applications produce high volumes of data that have only local relevance making storing or processing them remotely in the cloud an unnecessary burden on the network and cloud resources [3].

Efforts are made to tackle the issues of cloud computing from different perspectives. Mathematical optimization modelling is used as valuable tool to formulate alternative network solutions with the objective of energy minimization. The authors in [5-7] studied green and renewable energy resources in core networks. Analysis of big data and its impact on networks energy consumption is carried in [8-11]. Optical networks architecture designs and relative resilience and fault tolerance schemes are tackled in [12-18]. The authors in [4, 19-21] designed content distribution schemes for better resources utilization and improved energy efficiency. The authors of [22] modelled energy-efficient virtualization in cloud networks. Also, [23] introduced models for virtual machine (VM) placement in core networks. Paradigms, such as IoT [24-30] and Fog Computing [31, 32], benefiting from resources in end users proximity, are being actively investigated. The resilience of such paradigms is explored in [27, 33] ,which is an important measure of their reliability and quality of service as alternatives for cloud.

The term vehicular cloud [34] was coined to describe exploiting underutilized vehicular processing resources to deliver cloud services to end users. This approach usually goes hand in hand with the use of edge and cloud computing [35, 36] to complement the vehicles limited processing capacities, maintain quality of service, and facilitate system management. Research on vehicular networks has been continuously active. At early stages, the efforts were focused on benefiting from smart vehicle resources in transportation and safety related applications and services [37]. The concept of vehicular networks is still evolving, and research efforts are investigating different perspectives. In [38, 39] the energy efficiency and quality of service (QoS) were studied for different routing and base stations optimization schemes for vehicular networks in a city environment. The same authors in [40] and [41] developed position-based routing schemes for vehicular networks which they proved have better performance than flood-based protocols. In [42], the authors envisioned the use of resources available in vehicles parked in long term parking of airports and introduced the idea of using these resources as a datacenter. Also a two tier data center system in a paking lot was introduced in [43], in which the cost of communicating with conventional data centers was reduced by using storage resources of automotives in parking lots instead. In [44] the authors improved the energy efficiency of content distribution to city veicular users by using renewable energy and adaptive caching points. They also studied the impact of load adaptive caching points on the energy effencicy in [45], and several vehicular network scenarioes with energy efficeint adaptive/non-adaptive fog servers with renwable/ non-renwable energy in [46].

Our work in [47] introduced an energy efficient vehicular cloud architecture, to be used in distributed processing. In addition to vehicular processing, the architecture provided processing at edge nodes and the conventional cloud. Generic smart city applications were considered to evaluate the performance of the architecture. We evaluated scenarios of multiple requests of varied sizes and showed promising results of 70-90% power savings over conventional cloud for small-sized requests and 20-30% for medium and large requests. In this paper, more test cases are reported to further establish the merits of the energy efficient vehicular cloud architecture. The contributions of this paper can be summarized as follows:

(i) presenting for the first time the mixed integer linear programming (MILP) model developed to optimally allocate processing demands to the three layers of the architecture with the objective of minimizing the power consumption,
(ii) comparing the energy efficiency of processing scenarios considering different processing layers,
(iii) evaluating different test cases considering varying demand sizes, varying number of demands and the impact of processing demand splitting,
(iv) developing a heuristic to allocate processing demands in real time and comparing its performance to the MILP model.

The subsequent sections of the paper are organized as follows: The proposed architecture and the MILP model are introduced in Section II. The results of the model are presented and analyzed in Section III. The heuristic and its results are given in IV and the paper is concluded in V.

## II. PROPOSED VEHICULAR CLOUD ARCHITECTURE

### A. Architecture

In the following subsections we present the proposed energy efficient vehicular cloud architecture from three different perspectives: the processing layers, network communication interfaces, and control and coordination.

#### 1) Processing Layers

As seen in Figure 1, the proposed distributed architecture is composed of three processing layers:

**Vehicular Processing Layer:** The first layer is composed of vehicles equipped with high-performance on-board units. Other vehicles can serve the demand if they are willing to share their resources. The vehicles can dynamically cluster to form temporary clouds. Each vehicular cloud is formed under the control of an edge node.

**Edge Processing Layer:** The second layer is formed by edge nodes equipped with mini servers dedicated for smart city applications. This layer provides users with another nearby processing destination. In addition to the mini servers, the edge node encompasses an access point (AP) to communicate with vehicles and an optical network unit (ONU) to connect to the passive optical access network (PON).

**Cloud Processing Layer:** The last layer is the conventional cloud, which is geographically distant but has powerful computing capabilities. The cloud is connected to a core network node through switches and routers.

The MILP formulation introduced in the next section identifies the optimum solutions given the trade-off between having powerful energy efficient servers at far located clouds, accessed by traversing multiple network layers, and using the less powerful, less energy efficient closer processing resources offered by vehicles and edge nodes.

#### 2) Communication Interfaces

The use of vehicles provides communication technologies heterogeneity, which is both an attractive and challenging feature of vehicular networks. Vehicles support different communication interfaces including dedicated short-range communication (DSRC), bluetooth, WiFi, and cellular, and a lot of effort is dedicated to the optimal usage of these interfaces [48, 49]. In the proposed architecture, vehicles communicate with each other using DSRC as it provides high data rate and good coverage [50, 51]. The vehicle to vehicle (V2V) communication is not limited to the same vehicular cloud. Vehicles belonging to different clouds within the communication range of each other can communicate peer to peer using DSRC. For communication between vehicles and edge nodes (V2E), WiFi is used. Edge nodes communicate with each other in peer-to-peer manner using WiFi. Higher data rate V2V communication can also be supported through the WiFi interface. Edge nodes, as mentioned above, are equipped with an ONU to connect to higher layers through a PON access network. The inter-communication between the devices composing an edge node is through Ethernet of high speed and low energy per bit, so the power consumed is negligible.

#### 3) Control and Coordination

The decisions of where to serve a user demand, how much of this demand is to be served in a specific location, how the data associated with the processing demand is routed to that location, fall under the umbrella of system control. These decisions need to be optimized to reduce the power consumption.

Generally, the main challenge of the vehicular architecture is the dynamicity and variation of the resources. Keeping track of these changes is crucial to making educated decisions. As mentioned before, each vehicular cloud is controlled by an edge node. Each edge node is assumed to have knowledge of the vehicles under its control and Edge nodes exchange information about their vehicular clouds resource availability. Based on this knowledge each edge node takes decisions on where to process demands coming from its vehicular cloud. In this work, the overhead created by the control and coordinating data is not considered. In such a distributed control architecture the concept of software- defined networks (SDN) comes into play [52]. The architecture can be further enhanced with the addition of a layer of *centralized* controller with global view of the network, to oversee and coordinate the edge nodes. Whether using the distributed or centralized approach, the need for dynamic response and frequent updates on the architecture remain the same.

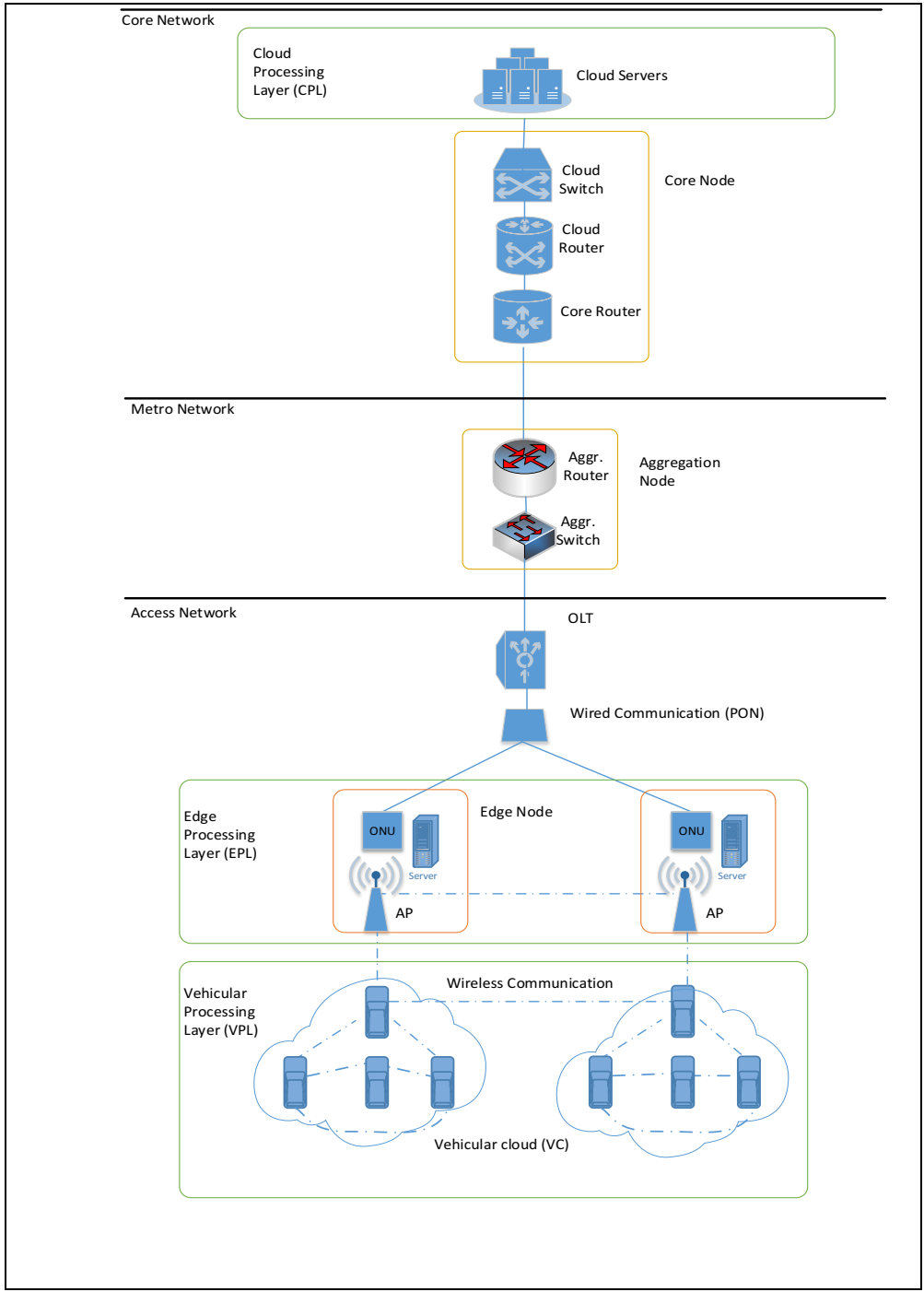

*Figure 1:End-To-End Vehicular Cloud Architecture*

### B. MILP Model

A MILP model is developed to optimize the selection of processing destinations used to place processing demands and the routing of traffic demands between source nodes and processing destinations in the proposed architecture while minimizing the power consumption, which is composed of processing power consumption and networking power consumption. Table 1 and Table 2 list the parameters and variables of the model.

*Table 1: Table of Model Parameters*

| | |
|---|---|
| $ND$ | Set of vehicles |
| $ED$ | Set of edge nodes |
| $SD$ | Set of cloud servers |
| $OLT$ | Set of OLT devices |
| $MD$ | Set of metro nodes |
| $CD$ | Set of core nodes |
| $N$ | Set of all nodes in the architecture |
| $Nm_n$ | Set of neighbouring nodes of node $n$, $n \in N$ |
| $U_s$ | Processing demand generated by node $s$ (MIPS), $s \in N$ |
| $V_s$ | Traffic demand generated by node $s$ (Mbps), $s \in N$ |
| $C_n$ | Processing capacity of node $n$ (MIPS), $n \in N$ |
| $K_n$ | Processing efficiency of node $n$ (W/MIPS), $n \in N$ |
| $S$ | Maximum number of processing nodes to process a demand |
| $B_{nm}$ | Maximum data rate on the link between $n$ and $m$ (Mbps), $n \in N, m \in Nm_n$ |
| $B_n$ | Maximum data rate node $n$ can support, $n \in N$ |
| $B^{VE}$ | Maximum data rate of the WiFi interface of a vehicle (this parameter is defined in addition to $B_n$ $n \in ND$ to account for the two communication interfaces of vehicles). |
| $B^{ONU}$ | Data rate of ONU at an edge node (this parameter is defined in addition to $B_n$ $n \in ED$ to account for the two communication interfaces of edge nodes). |
| $D_{nm}$ | Distance between node pair $n, m \in N$ |
| $NM_n$ | Maximum power consumption of networking at node $n \in N$ (W) |
| $NI_n$ | Idle power consumption of networking at node $n \in N$ (W) |
| $PM_n$ | Maximum power consumption of processing at node $n \in N$ (W) |
| $PI_n$ | Idle power consumption of processing at node $n \in N$ (W) |
| $NM^{ONU}$ | Maximum power consumption of ONU at an edge node |
| $NI^{ONU}$ | Idle power consumption of ONU at an edge node |
| $TX$ | The maximum transmission power consumption of wireless interface |
| $T_{nm}$ | Wireless transmission energy per bit over link $(n, m)$ where $n, m \in ND \cup ED$ |
| $RX$ | Receiver sensitivity of wireless interface |
| $R_{mn}$ | Wireless reception energy per bit at node n over link $(m, n)$ where $n, m \in ND \cup ED$ |
| $\epsilon$ | Power amplifier factor for wireless communication |
| $E_n$ | Energy per bit of networking at node $n$, $n \in OLT \cup MD \cup CD \cup ED$ |
| $PUE_n$ | Power usage effectiveness of node $n$ $n \in N$ |
| $A, M$ | large constants |

*Table 2: Table of Model Variables*

| | |
|---|---|
| $TP$ | Total power consumption of the architecture |
| $W_n$ | Total power consumption at node $n \in N$ |
| $WN_n$ | Networking power consumption at node $n \in N$ |
| $WP_n$ | Processing power consumption at node $n \in N$ |
| $\Omega_{sd}$ | The amount of processing demand of source $s$ served by processing node $d$, $s, d \in N$ |
| $F_{sd}$ | Traffic demand between source node $s$ and processing node $d$, $s, d \in N$ |
| $\lambda_{nm}^{sd}$ | The amount of traffic demand between source node $s$ and processing node $d$ traversing link $(n, m)$ where $s, d, n, m \in N$ |
| $\alpha_{sd}$ | $\alpha_{sd} = 1$ if demand of source node $s$ is served by processing destination $d$, otherwise $\alpha_{sd} = 0$. |
| $Q_s$ | Total number of processing nodes serving demand of source $s$, $s \in N$ |
| $\beta_n^{NET}$ | $\beta_n^{NET} = 1$ if node $n$ is used for networking, $n \in N$, otherwise $\beta_n^{NET} = 0$ |
| $\beta_n^{PR}$ | $\beta_n^{PR} = 1$ if node $n$ is used for processing, $n \in N$, otherwise $\beta_n^{PR} = 0$ |
| $\beta_n^{ONU}$ | $\beta_n^{ONU} = 1$ if ONU at node $n$ is used, $n \in ED$, otherwise $\beta_n^{ONU} = 0$ |

The objective of the model is to minimize the architecture power consumption by optimizing the allocation of processing resources and routing of traffic. The total power consumption (TP) of the architecture is given as:

$$TP = \sum_{n \in N} W_n \quad (1)$$

where

$$W_n = PUE_n(WN_n + WP_n) \quad \forall n \in N \quad (1)$$

Equation **Error! Reference source not found.** gives the total power consumption of each node in the architecture. The power consumption is composed of processing-induced part and networking-induced part. In addition, the impact of the power usage effectiveness (PUE) is accounted for at each node. The PUE is the ratio of the total power consumed by a networking or computing device or datacenter to the power consumed by the IT equipment only. It is an important measure of efficiency as modern computing and networking nodes require non-computing components for their operation, such as cooling and ventilation systems. An ideal PUE is equal to 1, which means all the power is consumed in performing IT operations. The processing and networking devices are assumed to follow a linear profile where the power consumption is composed of an idle power consumption which is the power consumed to activate the device and load dependent power consumption as seen in Figure 2. The load dependent power consumption is obtained by multiplying the device load by the energy per bit ($E_n$).

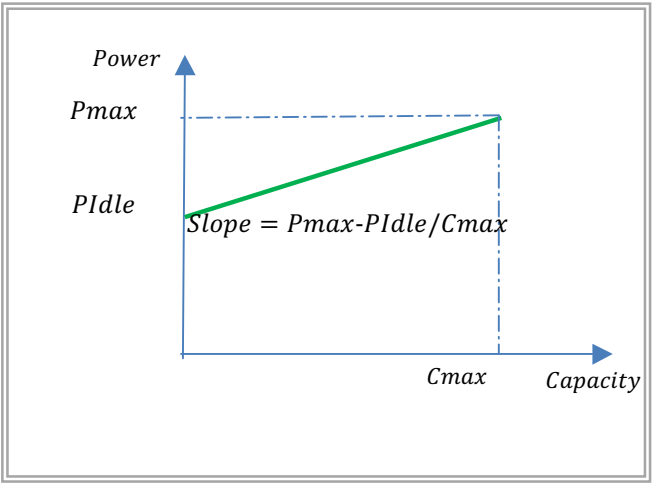

*Figure 2: Linear Power Profile*

In the following we give a detailed model of the network power consumption and processing power consumption of the different nodes in the network.

**For vehicular nodes:**

$$WN_n = \beta_n^{NET} \, NI_n + \sum_{s \in N} \sum_{d \in N} \sum_{m \in Nm_n} \lambda_{nm}^{sd} \, T_{nm} \; + \sum_{s \in N} \sum_{d \in N} \sum_{m \in Nm_n} \lambda_{mn}^{sd} \, R_{mn} \quad \forall \, n \in ND \quad (2)$$

Equation (2) gives the networking power consumption of a vehicular node as the sum of the OBU communication interface idle power consumption, the traffic-dependent, distance-dependent transmission power consumption, and the traffic-dependent reception power.

**For edge nodes:**

$$WN_n = \beta_n^{NET} NI_n + \sum_{s \in N} \sum_{d \in N} \sum_{m \in Nm_n \cap (ND \cup EB)} \lambda_{nm}^{sd} \, T_{nm} \; + \sum_{s \in N} \sum_{d \in N} \sum_{m \in Nm_n \cap (ND \cup EB)} \lambda_{mn}^{sd} \, R_{mn} + \beta^{ONU} NI^{ONU}$$
$$+ \sum_{s \in N} \sum_{d \in N} \sum_{m \in Nm_n \cap (OLT)} (\lambda_{nm}^{sd} + \lambda_{mn}^{sd}) E_n \quad \forall \, n \in ED \quad (3)$$

The edge node has two communication interfaces, the WiFi interface through its AP, and PON interface through the ONU. In equation (3), the first three terms calculate the AP (WiFi interface) power consumption, while the last two terms are for the PON interface power consumption. The PON interface power consumption is found by multiplying the traffic routed from the edge node ONU to the OLT by the energy per bit of the ONU. The idle power of the ONU is also added to the calculation.

**Wireless transmission and reception energy per bit:**

The traffic-dependent distance-dependent energy per bit for wireless transmission ($T_{nm}$) is given as

$$T_{nm} = \frac{TX}{B_{nm}} \; + \; \epsilon^* D_{nm}^2$$
$$\forall \, n \in ND \cup ED, m \in Nm_n \cap (ND \cup ED) \quad (5)$$

The first term of Equation (5) gives the traffic dependent part found by dividing the transmitter maximum power consumption ($TX$) by the link maximum data rate. The second term gives the distance-dependent power consumption as a function of transmission distance and the power amplifier factor.

The reception energy per bit for wireless transmission ($R_{mn}$) is found by dividing the node receiver sensitivity $RX$ by the link maximum data rate.

$$R_{mn} = \frac{RX}{B_{mn}}$$
$$\forall \, n \in ND \cup ED, m \in Nm_n \cap (ND \cup ED) \quad (6)$$

**The energy per bit is $E_n$ given as:**

$$E_n = \frac{(NM_n \text{-} NI_n)}{B_n}$$
$$\forall \, n \; \forall \, n \in OLT \, \cup MD \cup CD \quad (7)$$

$$E_n = \frac{(NM^{ONU} \text{-} NI^{ONU})}{B^{ONU}}$$
$$\forall \, n \in ED \quad (8)$$

**For OLT, metro, core nodes:**

$$WN_n = \beta_n^{NET} \, NI_n + \sum_{s \in N} \sum_{d \in N} \sum_{m \in Nm_n} \lambda_{mn}^{sd} \; E_n \quad (9)$$
$$\forall \, n \in OLT \cup MD \cup CD$$

The networking power consumption for the nodes from OLT to the core node is calculated by multiplying the networking energy per bit of each node, which is calculated in equations (7) - (8) by the traffic traversing it, as shown in equation (9).

**Processing power consumption:**

The processing power consumption at any node is given in equation (11) considering the processing idle power consumption and the processing load dependent power consumption which is a function of the node processing efficiency (the power consumed per MIPS as shown in equation (10)).

$$K_n = \frac{(PM_n - PI_n)}{C_n} \quad \forall \, n \in N \quad (10)$$

$$WP_n = \beta_n^{PR} PI_n + \sum_{s \in N} \Omega_{sn} \, K_n \, \forall \, n \in N \quad (11)$$

**The model is subject to the following constraints:**

$$U_s = \sum_{\substack{d \in N \\ d \neq s}} \Omega_{sd} \quad \forall \, s \in N \quad (12)$$

Constraint (12) states that the processing demand for a source node must be fully served by the processing destinations and the demand cannot be locally processed.

$$\sum_{\substack{s \in N \\ s \neq d}} \Omega_{sd} \leq C_d \quad \forall \, d \in N \quad (13)$$

$$\Omega_{sd} \geq \alpha_{sd} \quad \forall \, s, d \in N, s \neq d \quad (14)$$

$$\Omega_{sd} \leq A \, \alpha_{sd} \quad \forall \, s, d \in N, s \neq d \quad (15)$$

Constraint (13) ensures that the processing demands served by a processing node do not exceed its processing capacity. A binary variable is set to indicate that a node is selected as a processing destination for a demand source as shown in constraints (14) and (15).

$$F_{sd} = V_s \alpha_{sd} \quad \forall \, s, d \in N, s \neq d \quad (16)$$

The processing demand can be served in several nodes. Each one would receive the full traffic demand from the source, as stated by constraint (16).

$$\sum_{m \in Nm_n} \lambda_{nm}^{sd} - \sum_{m \in Nm_n} \lambda_{mn}^{sd} = \begin{cases} F_{sd} & if \, n = s \\ -F_{sd} & if \, n = d \\ 0 & otherwise \end{cases} \quad (17)$$

$$\forall \, s, d, n \in N, s \neq d$$

Constraint (17) is a flow conservation constraint. It ensures that the amount of traffic received by an intermediate node is equal to the amount re-transmitted. It also ensures that the traffic enters and leaves the node fully at the source and destination nodes respectively.

$$\sum_{s \in N} \sum_{d \in N} \sum_{m \in Nm_n} \lambda_{nm}^{sd} + \sum_{s \in N} \sum_{d \in N} \sum_{m \in Nm_n} \lambda_{mn}^{sd} \leq B_n \quad (18)$$

$$\forall \, n \in OLT \cup MD \cup CD$$

$$\sum_{s \in N} \sum_{d \in N} \sum_{m \in Nm_n \cap ND} \lambda_{nm}^{sd} + \sum_{s \in N} \sum_{d \in N} \sum_{m \in Nm_n \cap ND} \lambda_{mn}^{sd} \leq B_n$$

$$\forall \, n \in ND \quad (19)$$

$$\sum_{s \in N} \sum_{d \in N} \sum_{m \in Nm_n \cap ED} \lambda_{nm}^{sd} + \sum_{s \in N} \sum_{d \in N} \sum_{m \in Nm_n \cap ED} \lambda_{nm}^{sd} \leq B^{VE}$$

$$\forall \, n \in ND \quad (20)$$

$$\sum_{s \in N} \sum_{d \in N} \sum_{m \in Nm_n \cap (ND \cup ED)} \lambda_{nm}^{sd} + \sum_{s \in N} \sum_{d \in N} \sum_{m \in Nm_n \cap (ND \cup ED)} \lambda_{mn}^{sd} \leq B_n$$

$$\forall \, n \in ED \quad (21)$$

$$\sum_{s \in N} \sum_{d \in N} \sum_{m \in Nm_n \cap OLT} \lambda_{nm}^{sd} + \sum_{s \in N} \sum_{d \in N} \sum_{m \in Nm_n \cap OLT} \lambda_{mn}^{sd} \leq B^{ONU}$$

$$\forall \, n \in ED \quad (22)$$

Constraint (18) ensures that the maximum data rates of the OLT, metro, and core nodes is not exceeded. For vehicles, constraint (19) preserves the DSRC interface data rate (ie ensures that the interface data rate is not exceeded), which is used to communicate with vehicles. Similar constraints for the WiFi interface between vehicles and edge node are separately implemented in constraint (20), (21). Similarly, for the edge node when using the optical communication link through the ONU, the data rate is preserved through constraint (22).

$$Q_s = \sum_{\substack{d \in N \\ d \neq s}} \alpha_{sd} \qquad \forall \, s \in N \qquad (23)$$

$$Q_s \leq S \qquad \forall \, s \in N \qquad (24)$$

To benefit from the distributed processing resources, the division of the processing demand into smaller sub-tasks is allowed. Constraints (23) and (24) state the number of splits allowed.

$$\sum_{s \in N} \sum_{d \in N} \sum_{m \in Nm_n} \lambda_{nm}^{sd} + \sum_{s \in N} \sum_{m \in Nm_n} \lambda_{mn}^{sd} \geq \beta_n^{NET} \qquad \forall \, n \in N \qquad (25)$$

$$\sum_{s \in N} \sum_{d \in N} \sum_{m \in Nm_n} \lambda_{nm}^{sd} + \sum_{s \in N} \sum_{d \in N} \sum_{m \in Nm_n} \lambda_{mn}^{sd} \leq A\,\beta_n^{NET} \qquad \forall \, n \in N \qquad (26)$$

Equations (25) and (26) set a binary variable to 1 for nodes used in networking, i.e. transmit, receive, or relay nodes.

$$\sum_{\substack{s \in N \\ s \neq d}} \Omega_{sd} \leq \beta_d^{PR} \qquad \forall \, d \in N \qquad (27)$$

$$\sum_{\substack{s \in N \\ s \neq d}} \Omega_{sd} \geq A\,\beta_d^{PR} \qquad \forall \, d \in N \qquad (28)$$

Another binary variable is set to 1 in (27) and (28) to identify nodes used for processing.

$$\sum_{s \in N} \sum_{d \in N} \sum_{m \in Nm_n \cap OLT} \lambda_{nm}^{sd} + \sum_{s \in N} \sum_{d \in N} \sum_{m \in Nm_n \cap OLT} \lambda_{mn}^{sd} \geq \beta_n^{ONU} \qquad (29)$$
$$\forall \, n \in ED$$

$$\sum_{s \in N} \sum_{d \in N} \sum_{m \in Nm_n \cap OLT} \lambda_{nm}^{sd} + \sum_{s \in N} \sum_{d \in N} \sum_{m \in Nm_n \cap OLT} \lambda_{mn}^{sd} \leq A\,\beta_n^{ONU} \qquad (30)$$
$$\forall \, n \in N$$

A binary variable is set to 1 in (29) and (30) to indicate the use of the ONU in an edge.

## III. Evaluation and Results

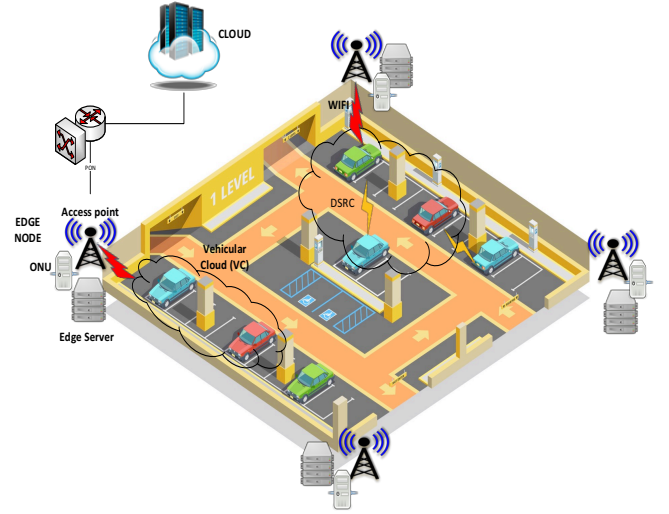

*Figure 3: Car Park Setting*

We evaluate the energy efficiency of the proposed architecture considering stationary vehicles in a parking lot. Vehicles in a parking lot offer resources for variable lengths of time, from short-term (half-hour to 3 hours) to long-term (over 3 hours to days and weeks) [53]. This accounts for and is due to vehicular mobility in and out of the car park. These resources are idle in congested business districts (e.g., employees' cars during working hours), or in urban areas with supermarkets and shopping malls, creating opportunities to exploit these resources in smart city applications.

Figure 3 illustrates a small parking area of 45 meters × 45 meters, accommodating up to 25 vehicles, with a standard parking space of 4.8 meters × 2.4 meters per car [54]. In our setting, we have 16 vehicles in the parking lot with the distance between two vehicles ranging from 2 meters to 24 meters. The parking lot is surrounded by 4 edge nodes, placed at an average distance of 30 meters away from the vehicles. Both DSRC and WiFi have communication ranges of several hundred meters [55, 56]. Each edge node serves a vehicular cloud of 4 vehicles.

The demands in this work are generated by the vehicles. However, the model is generic to accommodate having the demands produced by any node type, and this would provide further test cases for evaluation in future work. The demands cannot be locally processed, i.e. it is assumed that a vehicle cannot process its own demand (partly due to capacity constraints and partly to encourage cooperation with other vehicles). However, a vehicle having a demand can still process the demands of other vehicles. Our assumption is that for vehicular clouds to be applicable, cooperation of vehicles owners is required. This can be achieved if they are provided with the proper incentives to provide services to others while discouraging local service of their own jobs. The traffic demand (Mbps) and processing demand (MIPS) of a request are related based on the estimation in [57] for smart environment applications, which are summarized in Table 3 On average, one Mbps is sent for each 2000 MIPS of processing demand. Also, we choose the minimum size of traffic demand as 2 Mbps, which gives a processing demand of 4000 MIPS. The choice is made to allow for distributed processing among vehicles (vehicle processing capacity is 3200 MIPS in this study, see Table 4-4).

*Table 3: Traffic and Processing Requirements in Smart Environment Applications*

| Application Type | Average Data (kbps) | Estimated instructions/bit | Calculated Processing Requirement (MIPS) |
|---|---|---|---|
| City Monitoring Through WSN | 200 | 500-5000 | 100-1000 |
| Connected Vehicles | 200 | 50-2000-5000 | 10-400-1000 |

To calculate parameters for the vehicles as shown in Table 4, the following points were considered:

- Based on [58], highly efficient intel processors execute 4 instructions per cycle. Accordingly, 2 instructions per cycle per core are assumed for OBU in vehicles.
- From [59], OBU processor has 2 cores with Speed = 800 MHz which will be used to calculate the processing capacity in MIPS. Also, Maximum power is (OBUMAX = 10 W) and the idle power is (OBUI = 5 W).
- Based on [46, 60], a general-purpose computer spends 58% of its operational power on processing, 21% on storage (RAM and Disk), 21% on communication.
- So, for vehicles OBU, Processing Max/idle power = ((0.58 processing + 0.21 storage) × OBUMAX/OBUI).
- For networking, Networking Max/idle power = (0.21 × OBUMAX/OBUI)
- For WiFi, separate low power transceiver is added to the vehicle.
- For DSRC, modulation power (TX = +22 dBm [61] = 158 mW) and transceiver sensitivity (RX = -77 dBm [59]) values are used to find the energy per bit for transmission and reception, respectively.
- Similarly, for WiFi, modulation power (TX = +14 dBm [62] = 25 mW), and transceiver sensitivity (RX = - 72 dBm [62]) are used.
- For the vehicles PUE, the OBU is assumed to be small and efficient enough not to require ventilation system of any significant power consumption.

*Table 4: Vehicles Parameters Values*

| Notation | value |
|---|---|
| $C_n$ | 2 instructions/cycle x 2 cores x 800 MHz= 3200 MIPS [59] |
| $PM_n$ | OBU = 7.9 W [59] |
| $PI_n$ | OBU = 3.95 W [59] |
| $K_n$ | (7.9-3.95)/3200=0.00123 W/MIPS |
| $NM_n$ | 2.712 W [59], [62] |
| $NI_n$ | 1.05007 W [59] [62] |
| $B_n$ | 27 Mbps [61] |
| $B_{nm}$ | DSRC = 27 Mbps, WiFi = 150 Mbps |
| $B_n^{VE}$ | 150 Mbps [63] |
| $\epsilon$ | 100 pj/bit.m$^2$ [64] |
| $PUE_n$ | 1 |

To calculate parameters for the edge nodes as shown in Table 5, the following points were considered:

- As stated before, the edge node is composed of access point, server, and ONU, collocated in one place.
- For a server, a raspberry Pi processor is used.
- Based on [58], we assume 2 instructions per cycle for the raspberry Pi processor, with 4 cores of speed = 1200 MHz [65, 66].
- The power consumption of the raspberry Pi is dedicated for processing, and this assumption is used to find the processing efficiency.
- For the transmission and reception energy per bit, the transmit power (TX =28 dBm [63] = 630 mW) and reception sensitivity (RX = -104 dBm [63]) of the access point are used.
- The idle power of an edge node = sum of idle power of the three devices (Raspberry + AP + ONU). Using ON/Off power profile so only the component used is included the calculation.
- The three devices of the edge node are collocated in one place, but they are not contained or boxed, which provides natural cooling and ventilation and the PUE can be set to 1.
- The devices of the edge node are integrated together, so the inter-communication between them is ignored in this work.

*Table 5: Edge Nodes Parameters Values*

| Notation | value | |
|---|---|---|
| $C_n$ | Rasberry Pi | 2 instructions/cycle x 4 cores × 1200 MHz = 9600 MIPS [66] |
| $PM_n$ | Rasberry Pi | 12.5 W [66] |
| $PI_n$ | Rasberry Pi | 2 W [66] |

| $K_n$ | (12.5-2) /9600 = 0.0011 W/MIPS | |
|---|---|---|
| $NM_n$ | Access Point | 25 W [63] |
| $NI_n$ | Access Point | 5.5 W [63] |
| $B_n$ | 150 Mbps [63] | |
| $B_{nm}$ | 150 Mbps | |
| $NM_n^{ONU}$ | 15 W [67] | |
| $NI_n^{ONU}$ | 13.5 W [67] | |
| $B_n^{ONU}$ | 10 Gbps [67] | |
| $PUE_n$ | 1 | |

To calculate parameters for the cloud as shown in Table 6, the following points were considered:

- Based on [58], the server is assumed to run 4 instructions per cycle (highly efficient)
- The power consumption of the cloud is dedicated for processing, and this assumption is used to find the processing efficiency.
- Transmission power of cloud is ignored as it only receives data, the return of the result (small) to the source node is not considered.
- For the idle power consumption of cloud server, datasheets do not always provide specific number, and estimations are found when needed in the literature. According to [68], the idle power consumption of servers is about 60% of the maximum, but we are assuming a more efficient modern server which consumes an idle power of around 50% of the maximum.

*Table 6: Cloud Parameter Values*

| Notation | value |
|---|---|
| $C_n$ | 4 instructions/cycle x 10 cores x 2.8 GHz = 112000 MIPS [69] |
| $PM_n$ | 115 W |
| $PI_n$ | 57 W (50% of maximum) |
| $K_n$ | (115--57)/112000=0.000518 W/MIPS |
| $PUE_n$ | 1.1 [4] |

To calculate parameters for metro and core routers and switches, as shown in Table 7, the following points were considered:

- According to [70], machine to machine (M2M) traffic will be 7% of the global traffic by 2022.
- Connected cars traffic and connected cities, as part of the M2M traffic, are the fastest growing types of applications. Together they are assumed to make 13% of the traffic [70].
- So, for the portion of the idle power consumption of network devices attributed to our application, we are assuming $(0.07 \times 0.13)$ of the total idle power of each device. The idle power is taken as 90% of the maximum power, based on estimations in [71].
- The values for the PUE in the network devices (routers and switches) are derived from [4]
- For switches, it is assumed the devices become more power hungry as they get closer to the core and datacenter, since aggregation level and functionality become more complicated. Therefore, the aggregation switches are set to consume the typical power values in [72] and the cloud switches consume the typical power value stated in the datasheet [72].
- For the core and cloud routers, the port power consumption is estimated from [73], by diving the total power consumption (1450 W) by the number of ports (48 ports) as all ports have the same capacity.
- For the aggregation router, the port power consumption is estimated from [73], by dividing the maximum power consumption (420 W) by the maximum throughput (800 Gbps) to get the W/Gbps, which was then multiplied by the port capacity of 10 Gbps.
- The processing capacity of OLT, routers and switches in metro and core is set to zero.

*Table 7: Network Devices Parameters Values*

| Device Type | $NM_n$ | $NI_n$ (W) | $NI_n$ (%) | $B_n$ (Gbps) | $E_n$ ($\frac{nj}{bit}$) | $PUE_n$ |
|---|---|---|---|---|---|---|
| OLT [28, 74] | 1940 | 60 | 0.546 | 8600 | 0.219 | 1.5 [76] |
| Agg. Switch | 210 | 189 | 1.72 | 6*40 = 240 | 0.088 | 1.5 |
| Agg. Router port | 5.25 | 4.725 | 0.043 | 10 | 0.053 | 1.5 |
| Core Router port | 30 | 27 | 0.246 | 40 | 0.075 | 1.5 |
| Cloud router port | 30 | 27 | 0.246 | 40 | 0.075 | 1.5 |
| Cloud Switch | 470 | 423 | 3.85 | 6*100=600 | 0.078 | 1.5 |

We evaluate 4 scenarios of processing resources availability. In the first scenario requests can only be processed in vehicles

(V scenario). The second scenario optimizes the allocation of the processing request at vehicles and edge nodes only (VE scenario). In the third scenario, only conventional cloud has processing capacity (C scenario). This is the scenario used as benchmark for comparison. The last scenario optimizes the allocation of processing resources at the three processing layers (VEC scenario). We evaluate different test cases to study the impact of the traffic demand size, processing demand splitting and the competition over resources between multiple demands. We have test cases with only one vehicle generating a demand and other cases where we have demands generated by a number of vehicles. The number of splits allowed in the processing demand provide another evaluation perspective. As the processing demand is split, the associated traffic is delivered to each allocated destination. Also, the traffic associated with processing demand can be delivered fully or partially to the allocated processing node. An example of the full traffic delivery is a pedestrian collision avoidance application where one image is delivered to two processors and one of the processors is assigned to search for pedestrians for example, while the other processor searches for vehicles. An example of partial traffic delivery is sending half of the image to one processor and sending the other half of the image to the second processor where each processor searches for pedestrians in front of vehicles.

### A. Demand Size Variation

We consider a single vehicle to generate the demand. The processing demand is varied between 4000-60000 MIPS to reflect tasks with low processing requirements to high processing requirements. A processing demand can be split among any number of processing destinations, i.e. S is set to the maximum number of available processing nodes in the MILP model.

Figure 4 shows the processing demand placement in the three processing layers in each processing scenario. For the V scenario, the model saturates the vehicles in the same vehicular cloud of the vehicle generating the demand (VC1) before moving to other VC. In the VE and VEC scenarios, edge nodes are preferred because of their higher processing capacity and efficiency, so one edge node is packed for example, before activating more vehicular processing nodes as the demand grows larger in size.

Figure 5 shows the distribution of the architecture total power consumption between networking and processing. It clearly shows that processing is the dominating contributor to the power consumption. Figure 5 shows the merits of having the edge nodes as supporting processing resources. The VE scenario matches the optimal solution given by the VEC scenario, except for the last demand where the optimal solution is to serve the request in the cloud due to capacity limits of the vehicle and edge nodes. Further inspection of Figure 4.6 shows that for the V scenario, the networking power consumption started to increase significantly as the demand increases, until the total power exceeds the power consumed in the conventional cloud scenario. Also, it is worth mentioning that for the larger demands not served in the V and VE scenarios, the bottleneck is the networking capacity and not the processing capacity, e.g. for the largest traffic demand of 30 Mbps, the associated processing demand is 60000 MIPS while the total capacity of vehicles and edge nodes is 86400 MIPS. Therefore, increasing the bandwidth capacity can go a long way in improving the performance and power saving of the architecture.

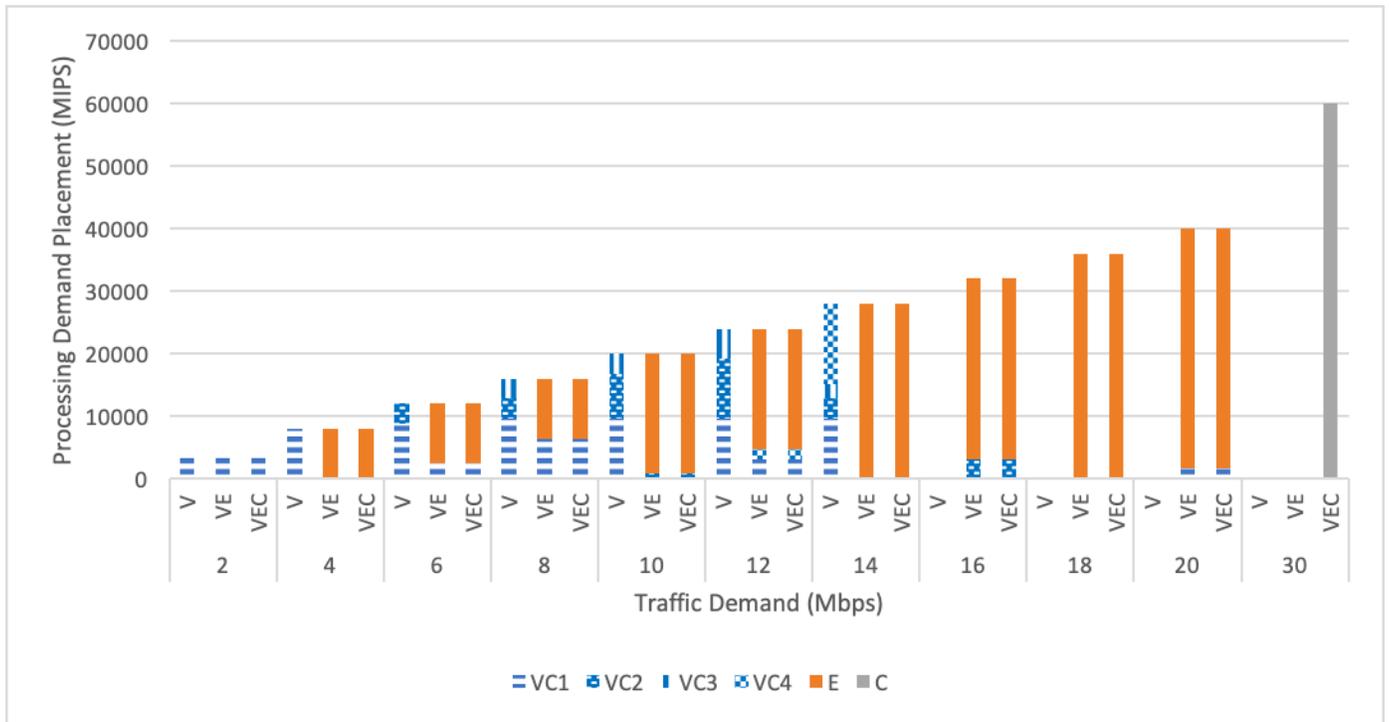

*Figure 4: Processing Demand placement when serving a single demand considering the different processing scenarios*

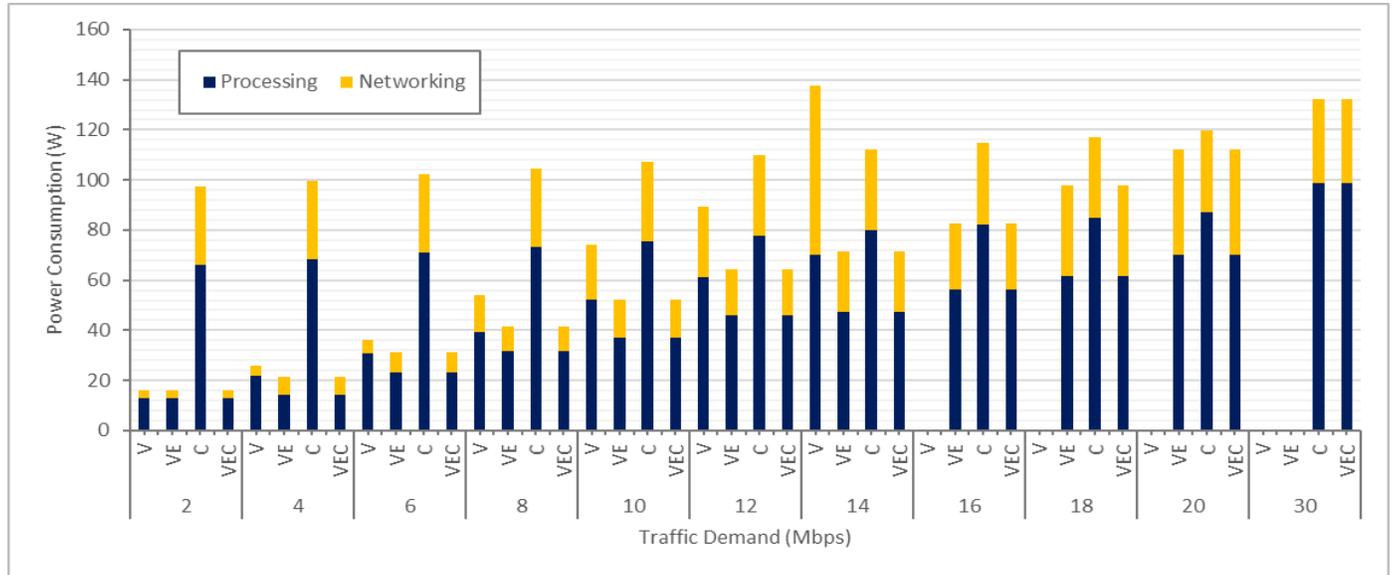

*Figure 5: Total power consumption when serving a single demand considering the different processing scenarios*

In Figure 6, we break down the networking power consumption. Figure 6 shows a surge in the networking power consumption of the V scenario serving a traffic demand of 8 Mbps. This is because 5 processing vehicles are needed to fully serve the demand (8 Mbps is associated with 16000 MIPS). This generates a total traffic of 40 Mbps to be transmitted from source to the processing destinations, which exceeds the capacity of the DSRC and necessitate the use of the WiFi and therefore leads to the increase seen in the power consumption. The other surge of the V scenario at 14 Mbps is due to edge nodes communicating through the PON access networks as the edge nodes WiFi APs cannot support this data rate.

Figure 7 breaks down the processing-induced power consumption. For the VEC scenario, the vehicle OBU are optimal as long as the total idle power of processing vehicles is lower than that of a single edge node. For example, at 2 Mbps (4000 MIPS), it is optimal to split the demand between two vehicles, while for the 4 Mbps demand (8000 MIPS), activating one edge server is more efficient than activating 3 vehicles OBUs. A combination of OBUs and edge node servers resulting in minimum power consumption is activated to serve higher processing demands.

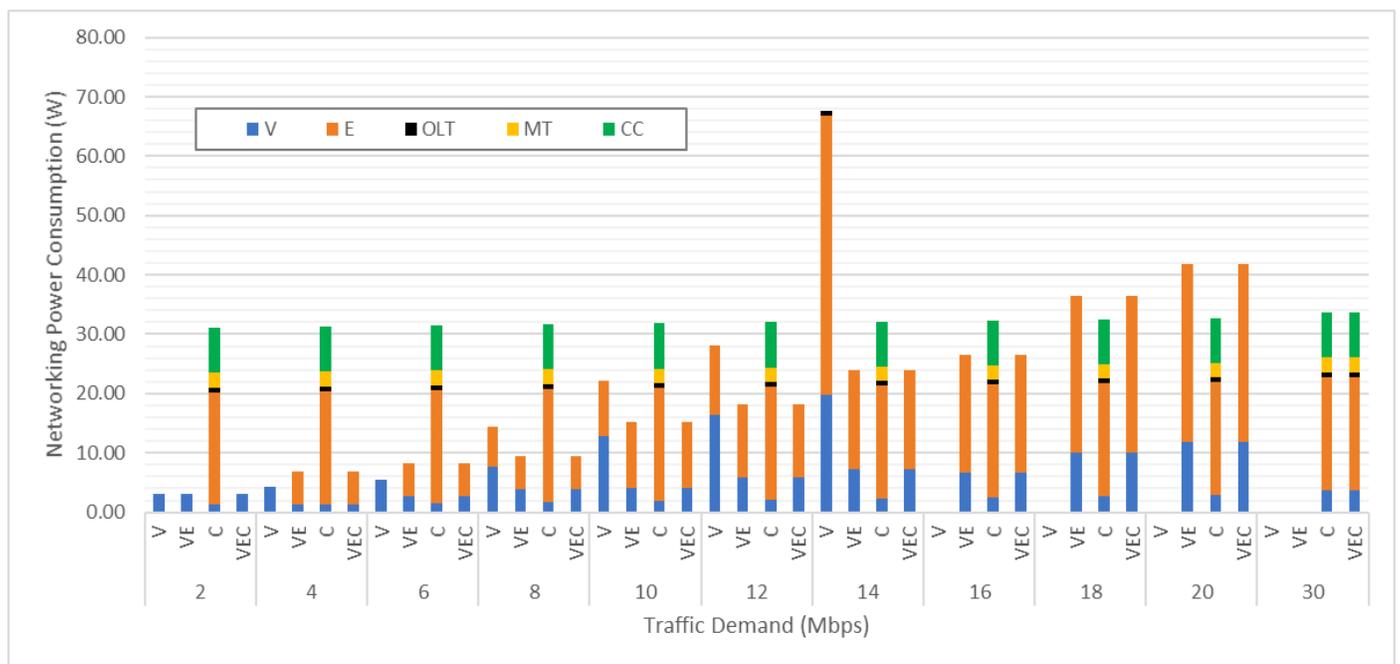

*Figure 6: Networking power consumption when serving a single demand considering the different processing scenarios*

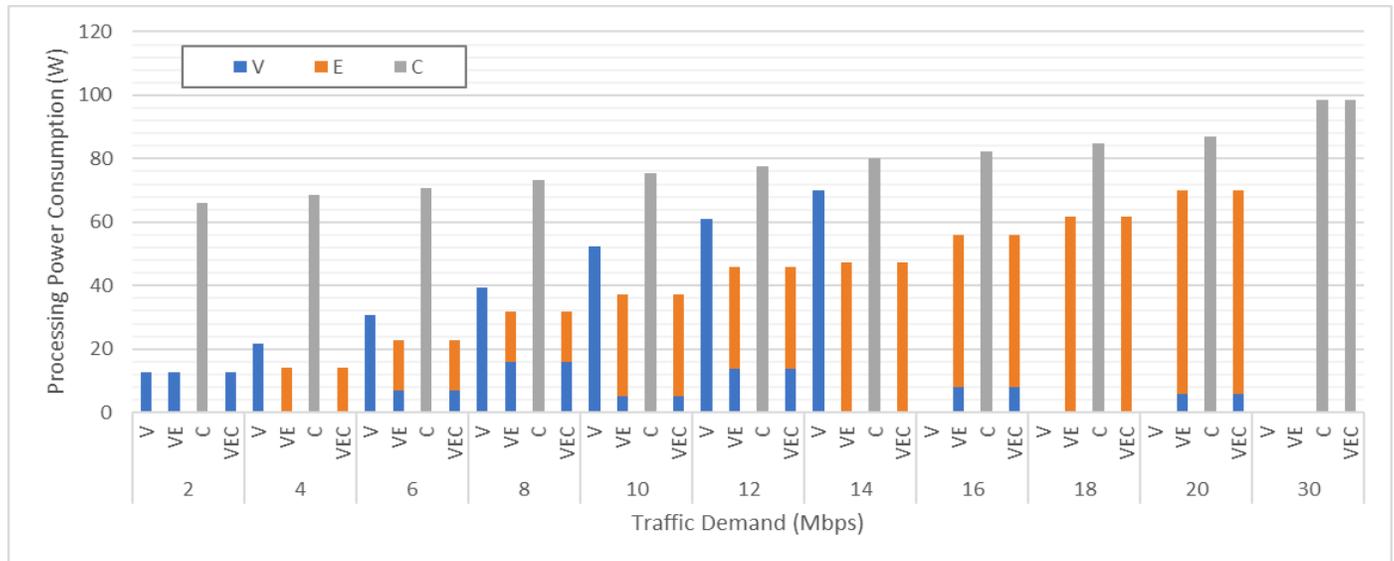

*Figure 7: Processing power consumption when serving a single demand considering the different processing scenarios*

Figure 8 shows the power savings of the three distributed processing scenarios (V, VE and VEC) in comparison with processing in the conventional cloud. Optimized processing in the VEC scenario resulted in power savings up to 84% compared to processing in the cloud.

Limiting processing to the vehicles and edge nodes (VE scenario) gave the maximum power savings for traffic demands as high as 20 Mbps. The energy efficiency of the vehicular only processing decreases as the size of the demand increases. Processing a demand of 14 Mbps in the vehicular cloud proves to be less efficient than cloud processing by 23%.

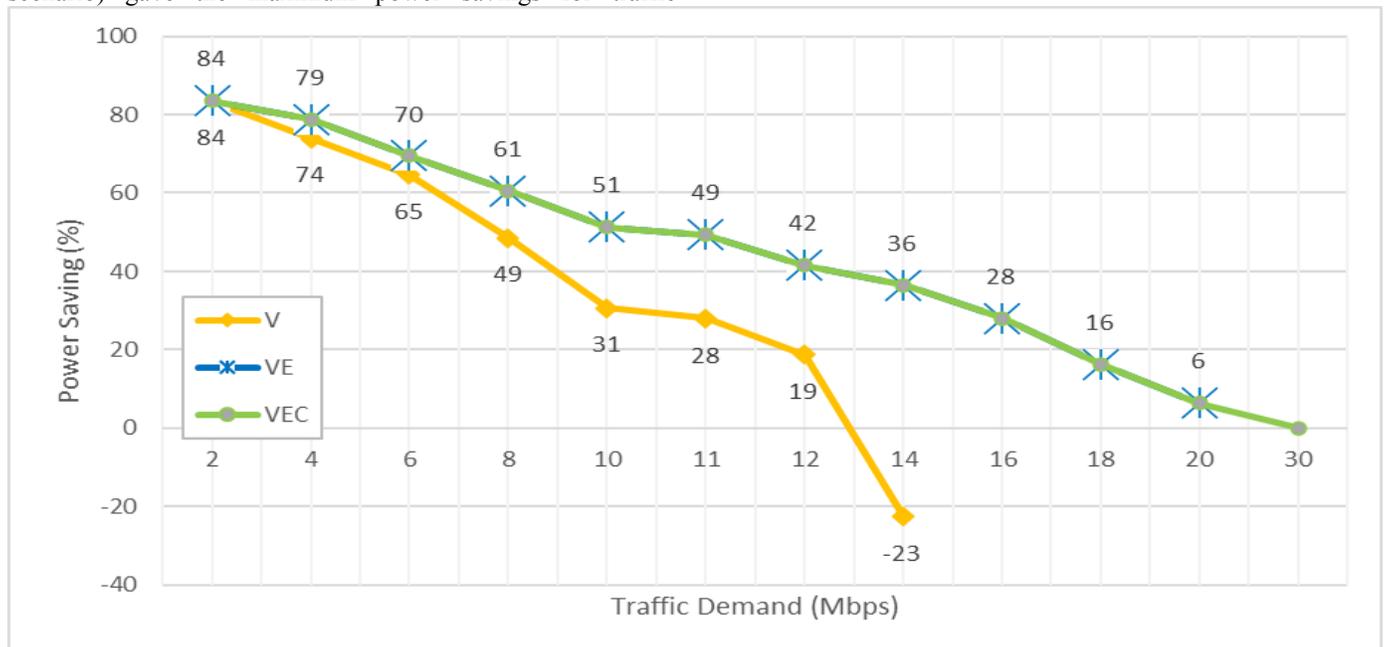

*Figure 8: Power saving of the three distributed processing scenarios (V, VE, VEC) in comparison with conventional cloud (single demand)*

### B. Processing Demand Splitting Limitation

Processing demands splitting can improve processing resources utilisation and consequently the energy efficiency of serving demands. It can also reduce the total processing time and avoid exhaustion of computational resources. In this subsection we study the impact of limiting the number of processing nodes that can serve the request as opposed to the unlimited processing demand splits studied in the previous subsection.

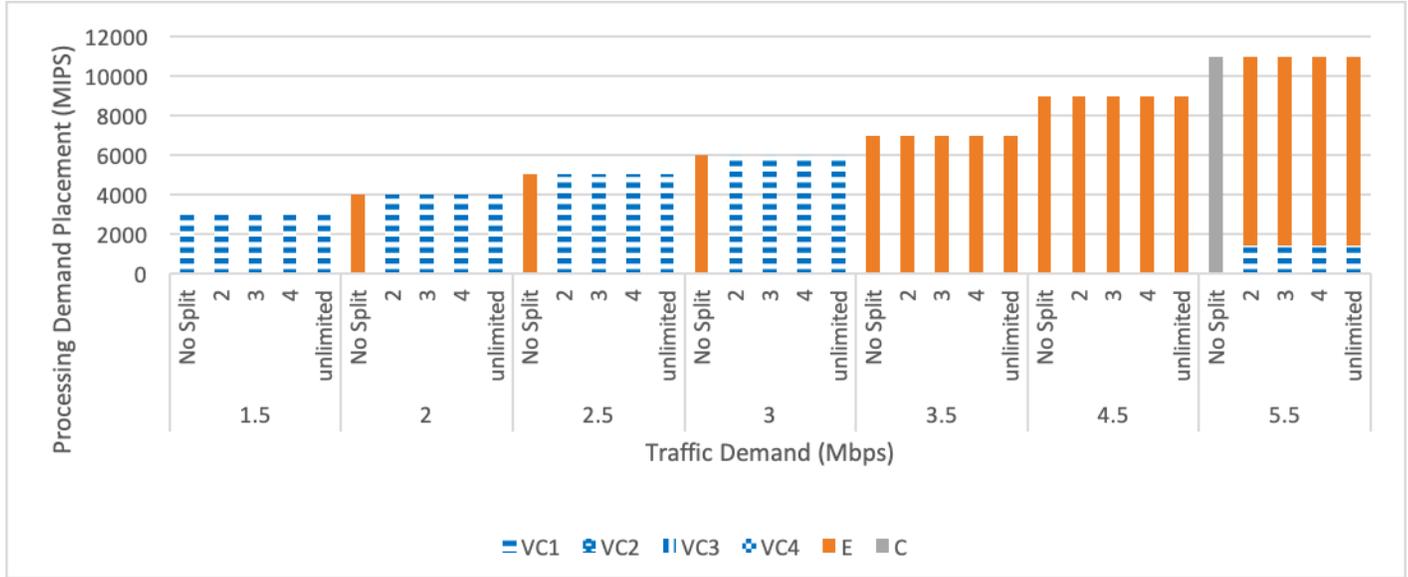

*Figure 9: Processing demand placement of the VEC scenarios with varying splits limits*

Figure 9 shows the processing placement and it shows only the use of one VC for small demand size and the ability to split.

Figure 10 shows the total power consumption of the VEC scenario under different splitting limits. It shows that a splitting limit of 2 is enough to achieve the minimum power consumption in this case. Splitting a request of 5.5 Mbps traffic demand between two processing destinations in a VEC scenario improves the energy efficiency by 71% compared to processing without splitting which results in processing the request in the cloud as seen in Figure 11.

From Figure 11, we can also infer that the usability of V scenario is limited when no splitting was allowed. Splitting traffic demands of 2-3 Mbps allows them to be processed in the vehicular cloud which slightly increases the processing power consumption, as seen in Figure 11, due to the lower processing efficiency of the OBUs. On the other hand, avoiding communicating with the edge nodes by processing in the vehicular cloud significantly reduces the networking power consumption as seen in Figure 12 although traffic is replicated to the two vehicular processing destinations serving the demand. The overall power reduction from the no splits case is 2-3%.

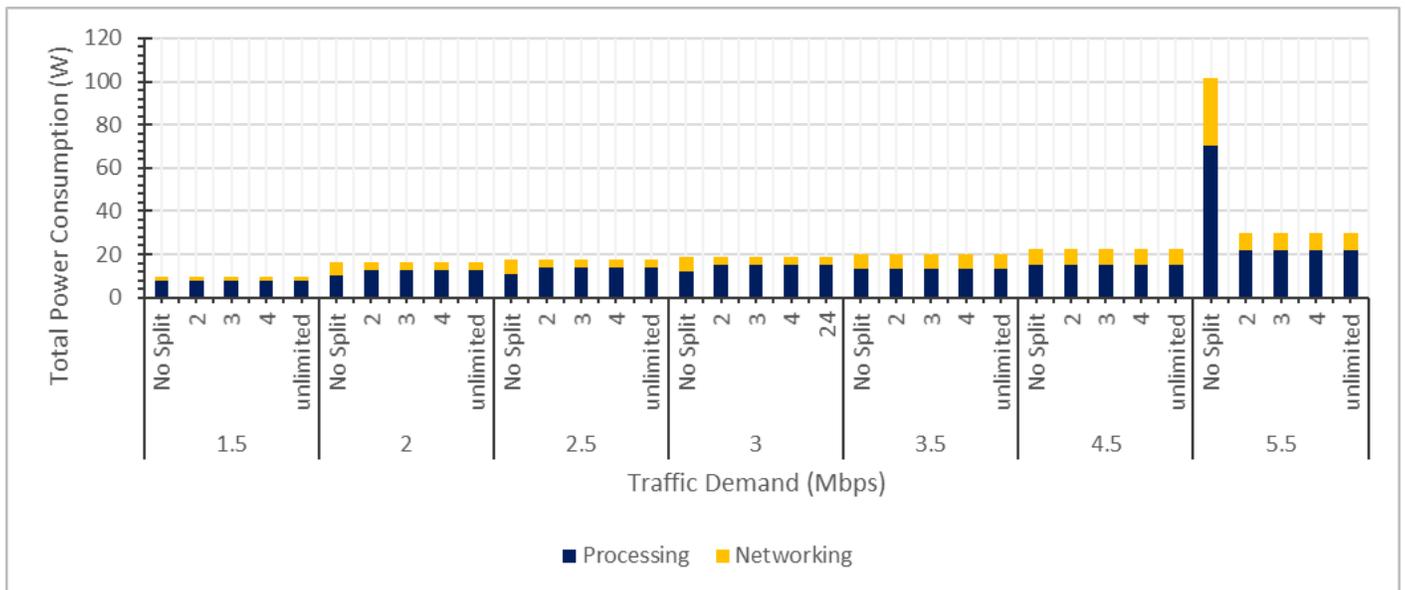

*Figure 10: Total power consumption of the VEC scenarios with varying splits limits*

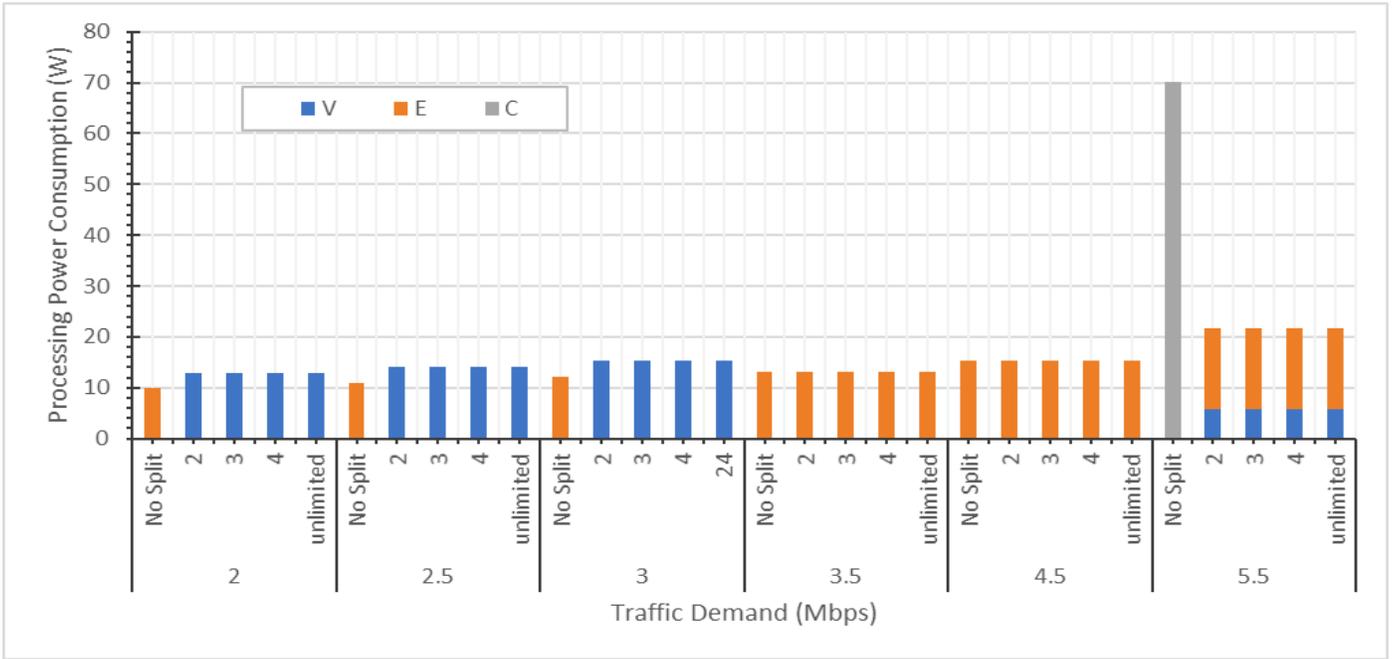

*Figure 11: Processing power consumption of the VEC scenarios with varying splits limits*

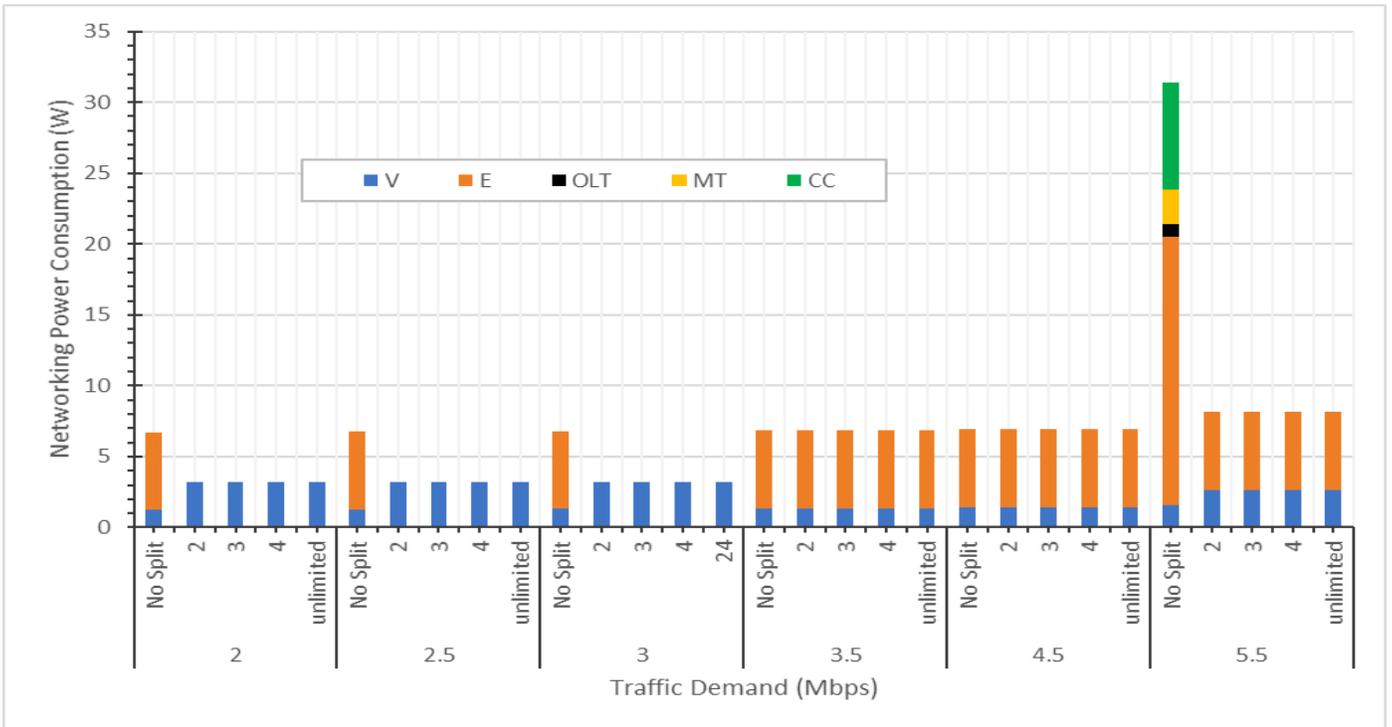

*Figure 12: Networking power consumption of the VEC scenarios with varying splits limits*

## C. Proportional Traffic Assignment

In all the previous subsections, we assumed that all processing destinations receive the traffic demand in full, even when serving part of the processing demand. This limits the efficiency of processing demands splitting as it burdens the network. However, for some types of applications, a processing node serving part of the processing demand requires access to only part of the data to be processed. An example of this can be an application processing multiple images or multiple videos as discussed earlier. In this subsection we optimize the processing of a demand with traffic that can be proportionally split among the nodes serving the demand, referred to as proportional traffic (PT) demand. This test case is compared to the one where the full traffic (FT) demand is delivered to every processing destination.

The MILP model is updated to represent PT requests, by replacing constrain (16) with the following equation:

$$F_{sd} = V_s \frac{\Omega_{sd}}{U_s} \qquad \forall s, d \in N, s \neq d \quad (31)$$

Equation (31) ensures that the traffic delivered to a processing node from a source node is proportional to the processing carried by the processing node.

The comparison considers a single demand of varying sizes for the 3 scenarios (V, VE, VEC), the C scenario is unaffected by this change as the whole demand is sent to the cloud.

Proportionally splitting the traffic relieves the V scenario traffic bottleneck observed for FT for traffic demands higher than 14 Mbps in subsection A. Comparison to Figure 5 for serving with full traffic (FT) shows that some cases in the V scenario that were infeasible before due to networking capacity limitations, are served in this modified case.

Figure 13 shows the processing placement in each layer and it shows that in the FT case, the V scenario packed one VC before moving to the next one, and it was able to serve bigger demands.

Figure 14 and Figure 15 compare the FT and PT cases under the V scenario in terms of processing and networking, respectively. Proportionally splitting the traffic among the processing destinations has not changed the number of processing destinations compared to the FT case. However, the networking power consumption was reduced. It improved the utilization of the DSRC communication bandwidth and therefore reduced the networking power consumption as the WiFi interface and the PON network are used less. Proportionally splitting traffic also relieves the traffic bottleneck observed for FT case for traffic demands higher than 14 Mbps. Similar trends are observed for the VE scenario as shown in Figure 16 and Figure 17. Improved utilization of the network bandwidth under the PT case allowed the VE scenario to serve 30 Mbps demands.

Comparing the PT and FT cases under the VEC scenarios in Figure 18 and Figure 19 confirms that the cloud is the optimal processing destination for the 30 Mbps demand even when the demand traffic resulting from distributed processing in the vehicular and edge layers can be supported by the network. This is due to the energy efficiency of the cloud in processing such a large demand compared to processing in the vehicular cloud and edge nodes which requires activating multiple processing destinations.

Proportional traffic impacts the power savings achieved by the different processing scenarios compared to processing in the conventional cloud, as seen in Figure 20. Proportionally splitting the traffic improves the energy efficiency of the V scenario compared to the FT case making processing demands as high as 20 Mbps in the vehicular cloud more efficient than cloud processing. As mentioned above, under the PT case the VE scenario can process demands as high as 30 Mbps. This is, however, less efficient than processing in the cloud by 14%. For the VEC scenario, the power savings improved from 6% for the FT case to 19% for the PT case for a demand of 20 Mbps compared to processing in the cloud.

.

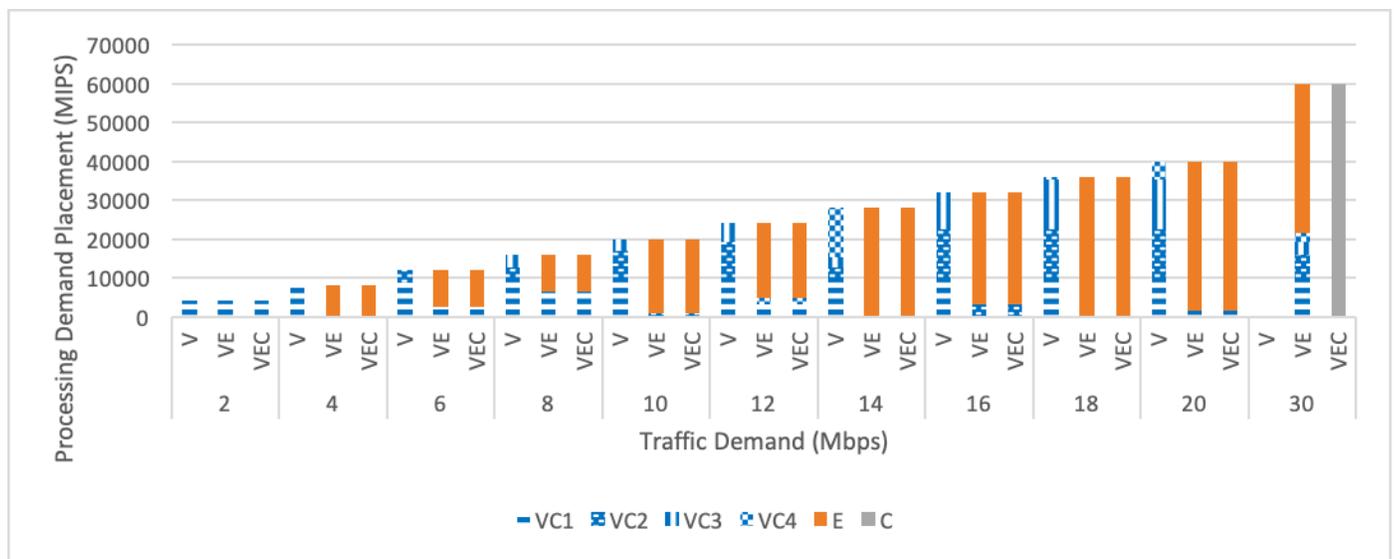

*Figure 13: Processing demand placement considering the three processing scenarios (V, VE, VEC) with proportional traffic (PT)*

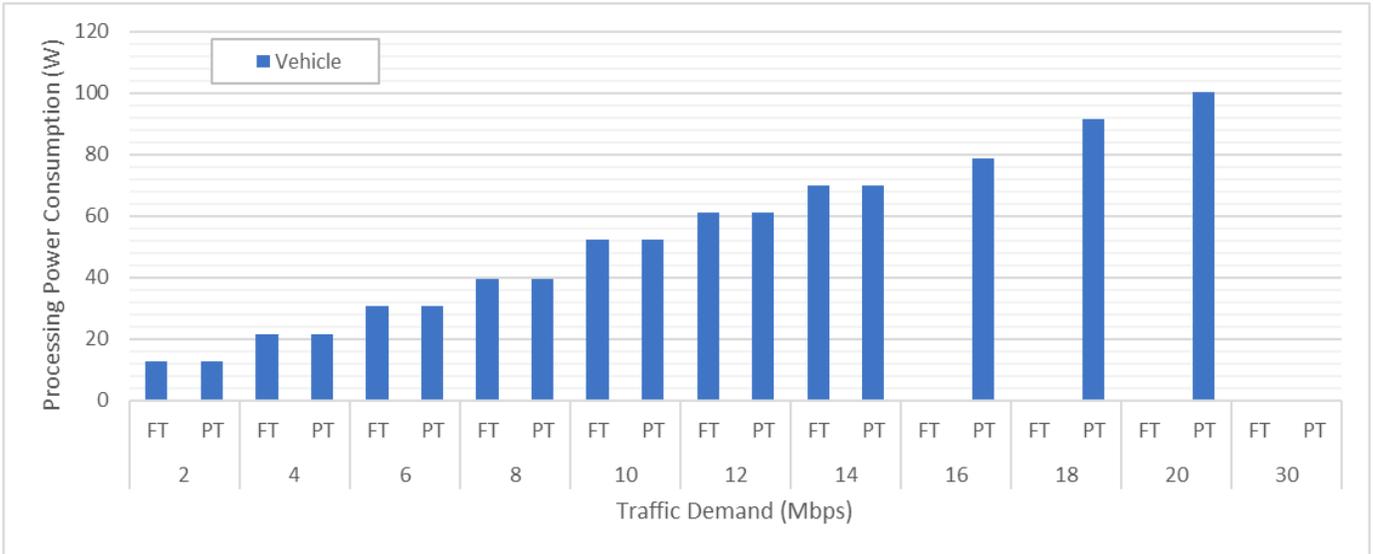

*Figure 14: Processing power consumption of the V scenario considering full traffic (FT) and proportional traffic (PT)*

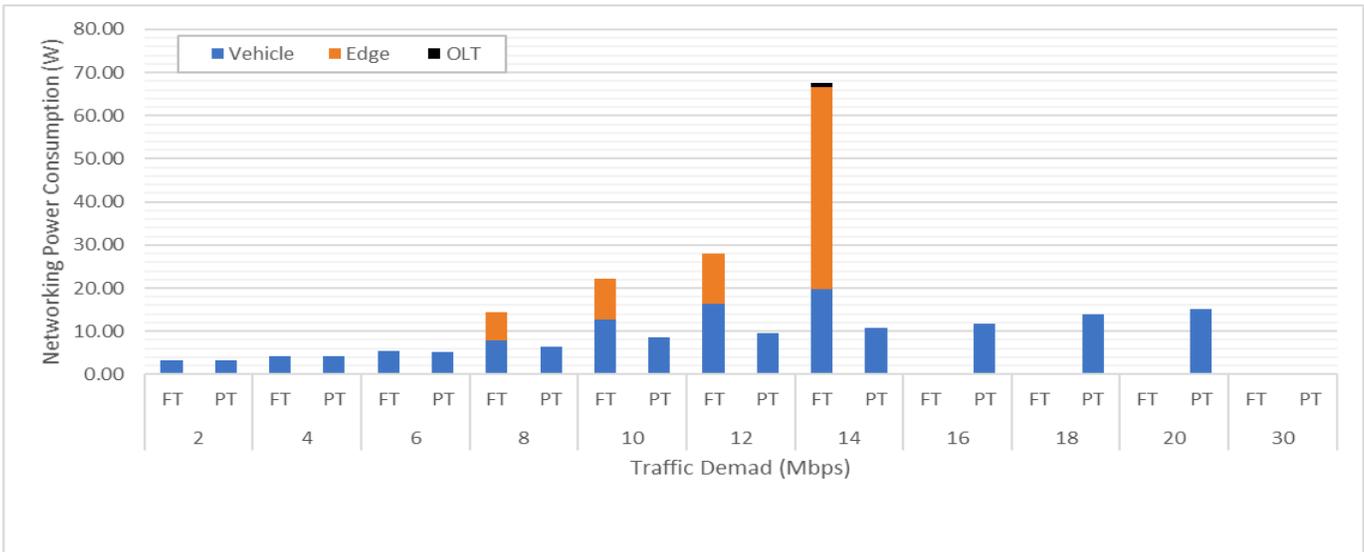

*Figure 15: Networking power consumption of the V scenario considering full traffic (FT) and proportional traffic (PT)*

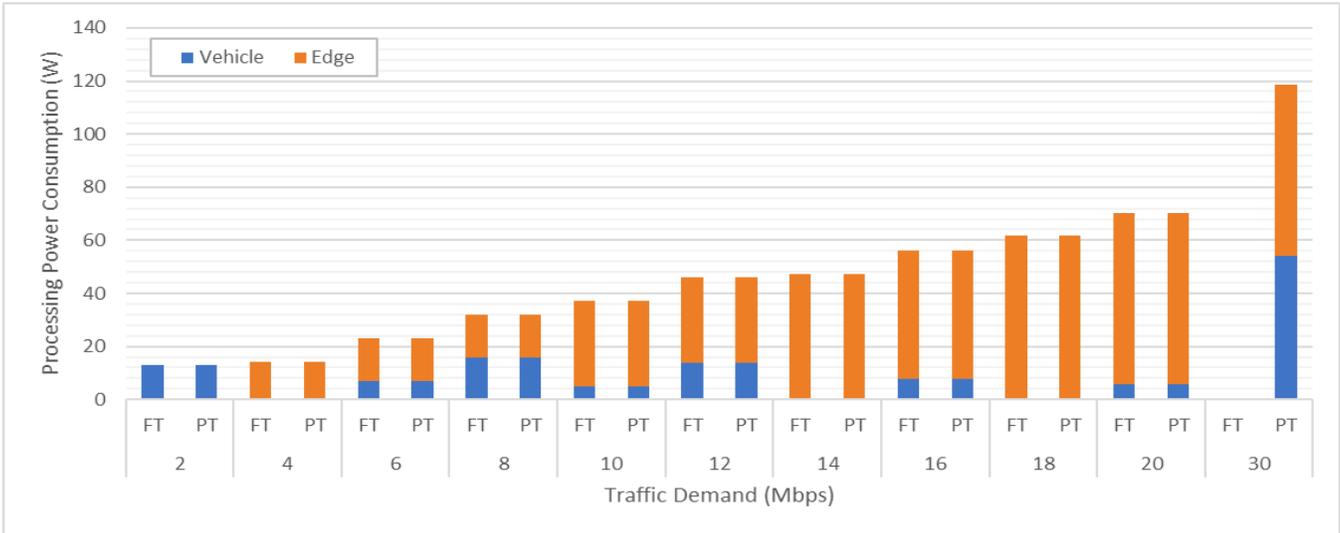

Figure 16: Processing power consumption of the VE scenario considering full traffic (FT) and proportional traffic (PT)

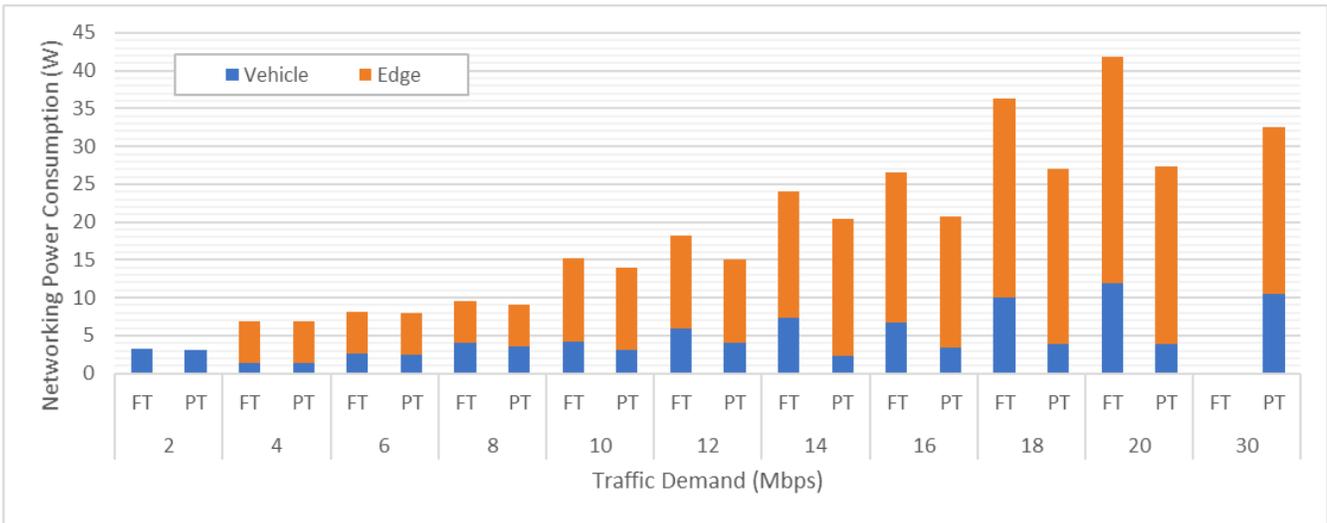

Figure 17: Networking power consumption of the VE scenario considering full traffic (FT) and proportional traffic (PT)

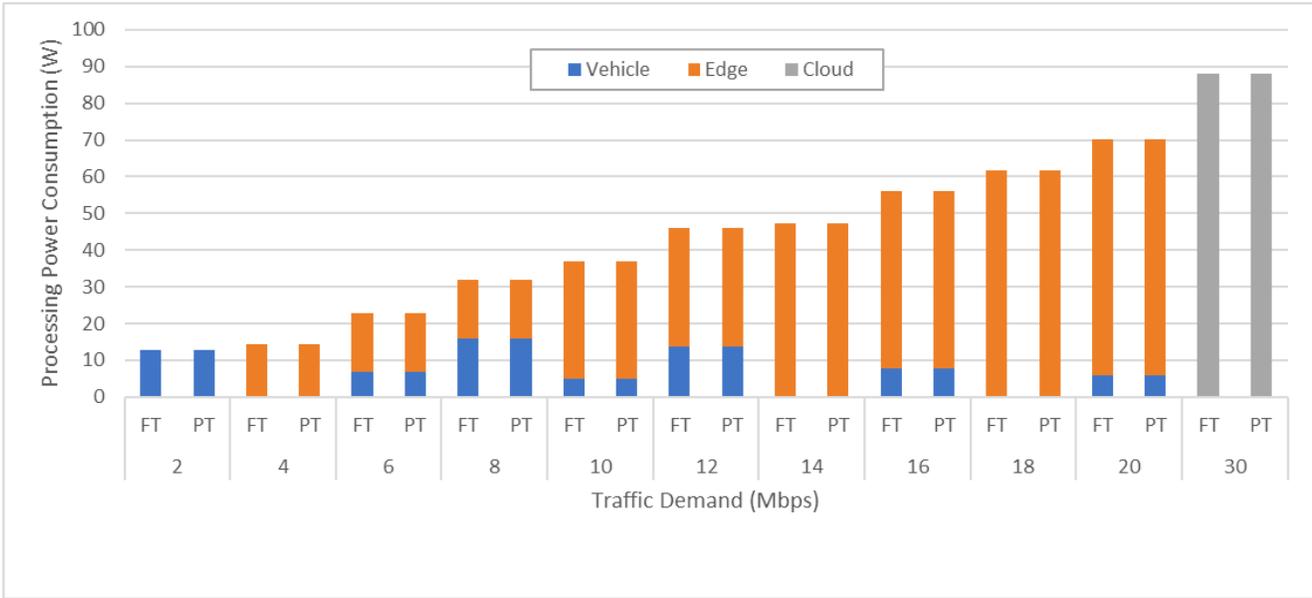

*Figure 18: Processing power consumption of the VEC scenario considering full traffic (FT) and proportional traffic (PT)*

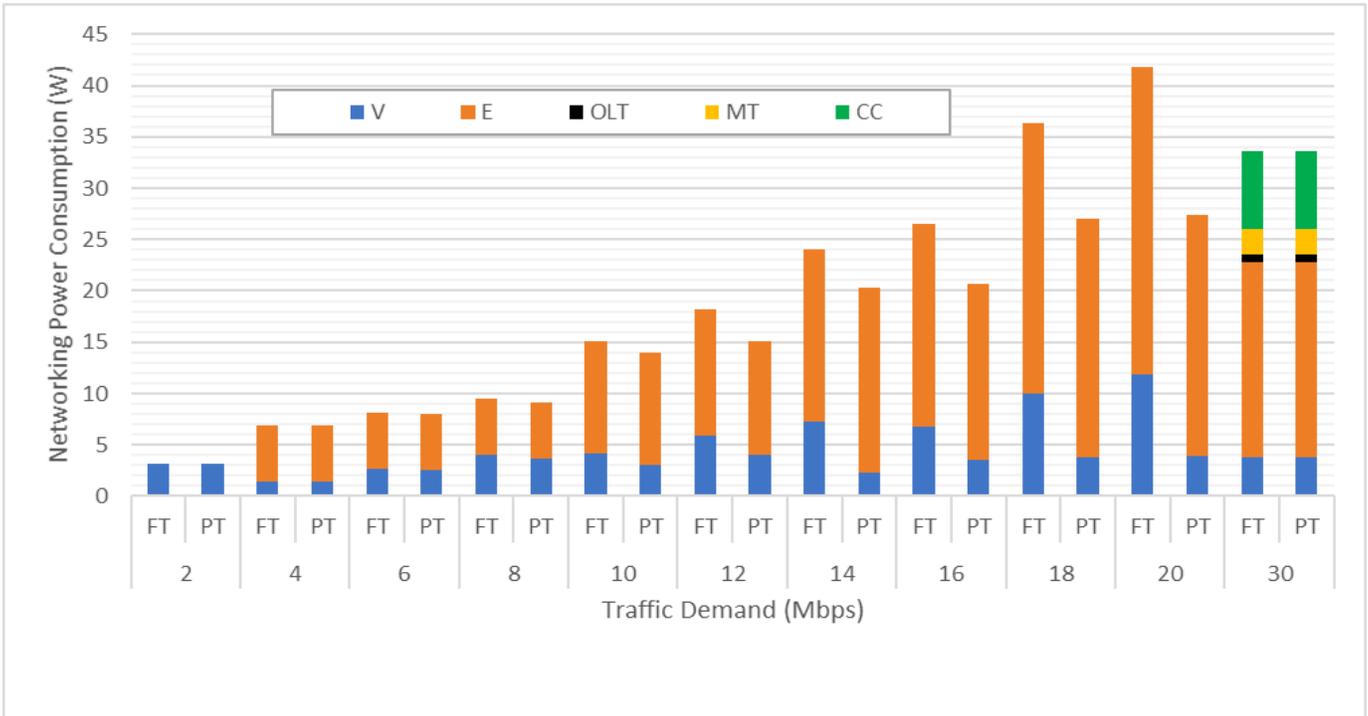

*Figure 19: Networking power consumption of the VEC scenario considering full traffic (FT) and proportional traffic (PT)*

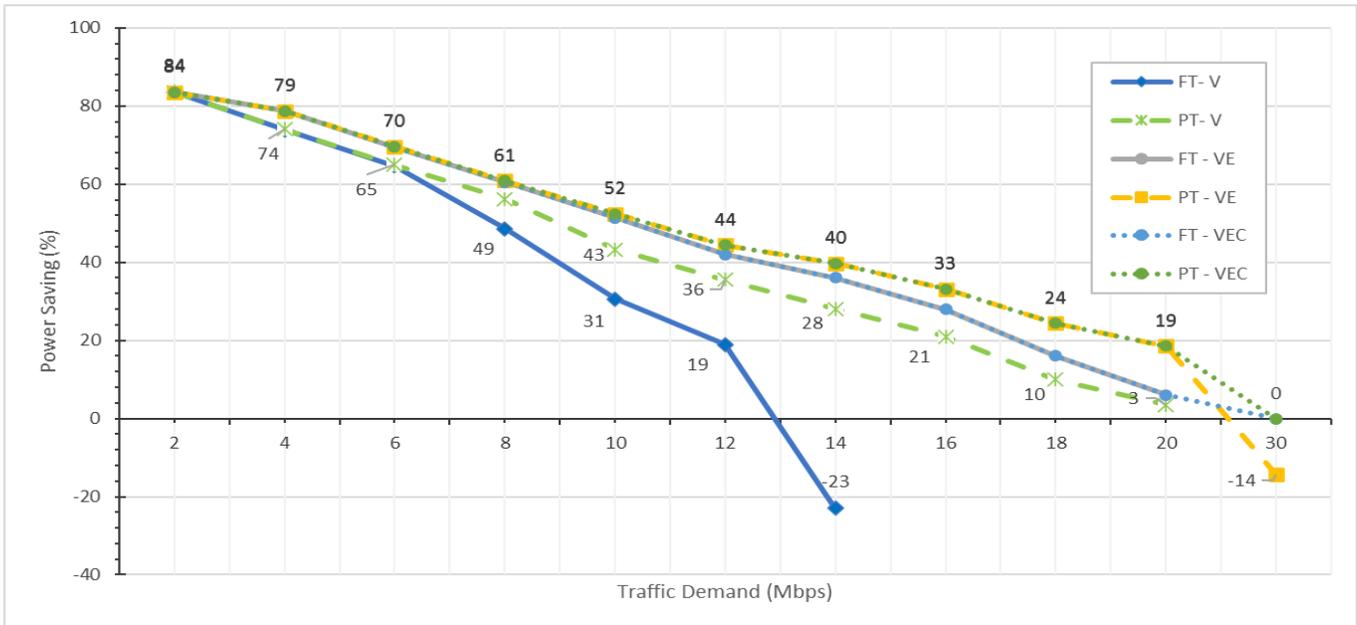

*Figure 20: Power saving of the (V, VE, VEC) FT and PT cases in comparison to the C scenario*

### D. Multiple Demand Service

In this section we examine multiple requests competing for the available resources under the VEC scenario considering full traffic replication to all processing destinations. Relative to the processing resources in vehicles and edge nodes, three demand profiles are examined: low (Traffic 1 Mbps, Processing 2000 MIPS), medium (Traffic 3 Mbps, Processing 6000 MIPS), and high (Traffic 5 Mbps, Processing 10000 MIPS).

The low demand can be served in single vehicle, medium demand can be served in single edge node, and the high demand exceeds the capacity of a single edge node. Figure 21 shows the processing demand placements as the demands grow in size/number, and it shows a tendency to use edge nodes over vehicles. Figure 22 evaluates the total power consumption of up to 10 co-existing requests and illustrates the impact of varying demand size/number on the processing and networking power consumption. With the growing requirements, it becomes less costly to operate the cloud as the difference between the high idle power consumption of the cloud server and the idle power consumption of the multiple vehicles and edge nodes required to serve the demands decreases. Also, for higher demand sizes the increase in the networking requirements cannot be accommodated with the use of the vehicles and edge only. The power savings of the VEC scenario rapidly decreases as the demands increase in size/number, as shown in Figure 23. For medium and high demands as the increase in the number of demands forced the use of the cloud, the power savings drops to 0%.

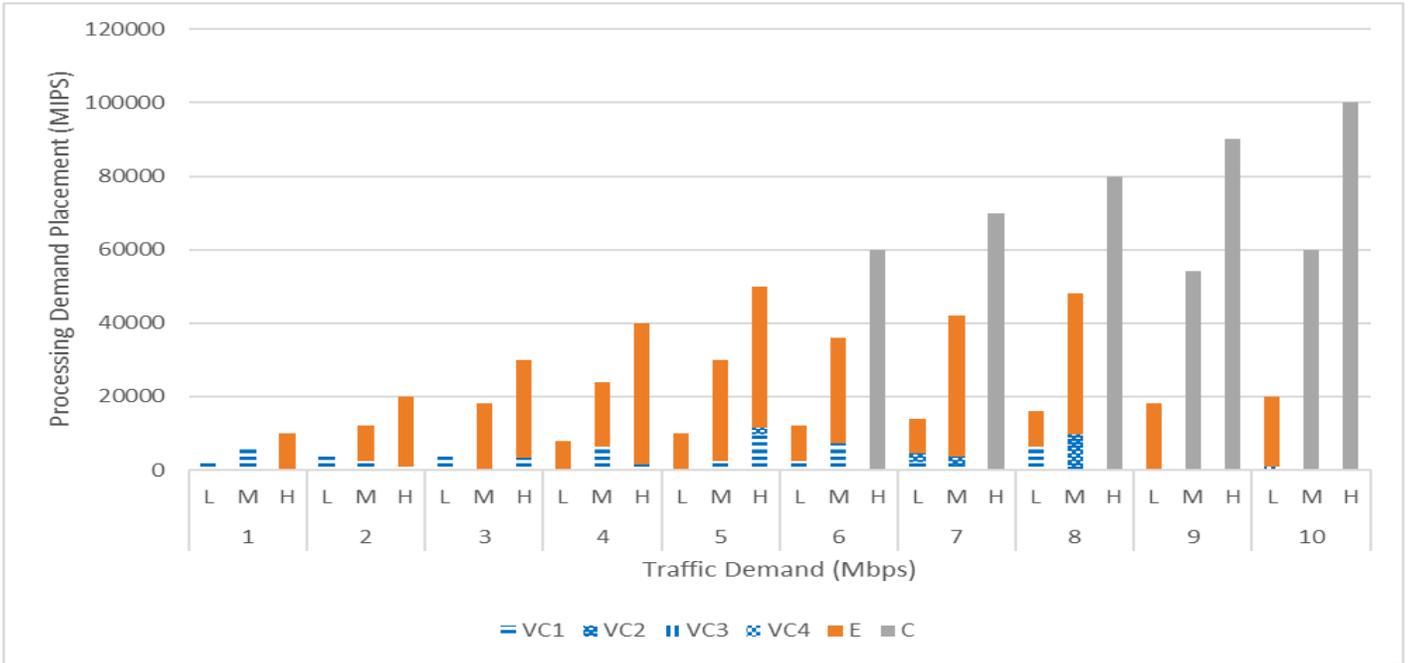

*Figure 21: Processing demand placement when serving multiple demands of varying sizes considering the VEC processing scenario*

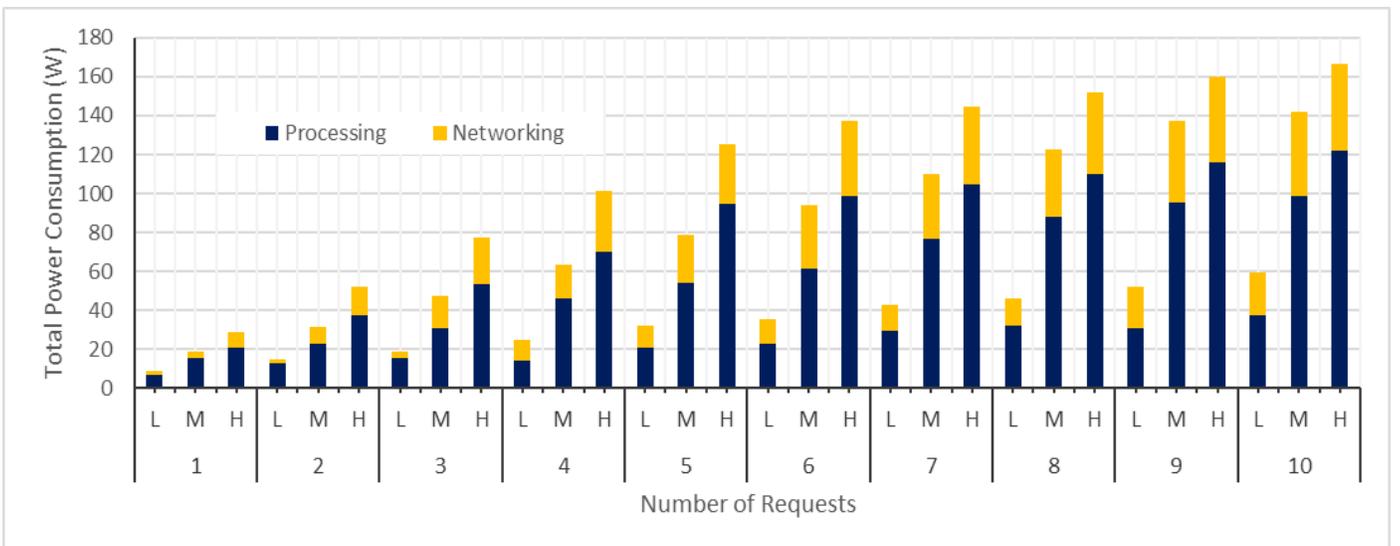

*Figure 22: Total power consumption when serving multiple demands of varying sizes considering the VEC processing scenario*

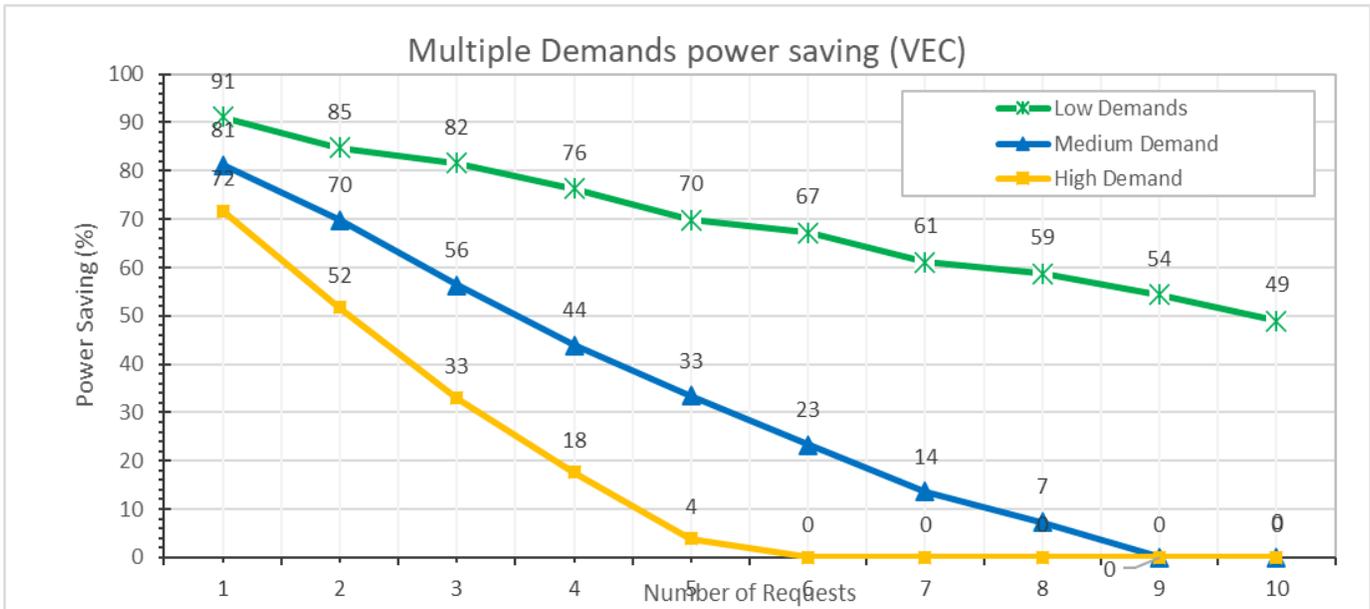

*Figure 23: Power Saving when serving multiple demands of varying sizes considering the VEC processing scenario*

## IV. ENERGY EFFICIENT DEMAND ALLOCATION HEURISTIC IN A VEHICULAR CLOUD ARCHITECTURE

A heuristic is developed based on insights obtained from the model to allocate processing demands in real time. The heuristic flowchart is shown in Figure 24. The heuristic serves demands in descending order of their processing requirements as the processing power consumption dominates the power consumed to serve demands. For each demand, the processing nodes are sorted based on the criteria defined in equation (32), with the *most fit* candidate in the beginning of the list and the *least fit* at the end. Sorting of the processing nodes can be ascending or descending depending on the demand size in comparison with the capacities of the distributed processing nodes. Also, the counter (*Trial*) is set to 1. This variable is set to give each demand two attempts to find a processing destination(s) by going over the complete list of processing nodes candidates and trying to route over them.

$$SortingCriteria_{sd} = NPower_{sd} + PrPower_{sd} + Idle_d \quad (32)$$
$$\forall\, s, d\, \in N$$

where:

$NPower_{sd}$ is the power consumption of routing traffic between source $s$ and processing destination $d$ over the minimum hops route.

$PrPower_{sd}$ is the processing power consumption of serving processing demand of source node $s$ in processing destination $d$

The heuristic selects the most fit candidate from the sorted list and tries to route the traffic demand over the minimum hop route if the candidate has available processing resources. If the minimum hop is not available due to networking capacity limitations, the heuristic removes the link of limited capacity from valid routes. The heuristics then selects the next node in the sorted list and tries to route the traffic demand on the minimum hop route. The heuristic examines all the nodes in the sorted list until all the demand under consideration is served. If all nodes are examined but the demand is not fully served, the trial counter is incremented by 1 and the heuristic examines nodes in the sorted list again. Note that the availability of the minimum hop routes will change in the second attempt, giving the processing nodes that were skipped before due to networking limitation a new chance in terms of the least hops route, and the processing node might be used to serve. Each demand is allowed only two attempts (more attempts can be allowed, at the cost of increased complexity) of examining the nodes in the ordered list to select a processing destination(s). Also, the number of processing nodes used is counted to ensure they do not exceed the allowed split limitation. Furthermore, a processing node is packed before moving to the next node.

If the demand is not served after the two attempts, it will be blocked and all the resources that were used to partially serve it are released to be used by other demands. The heuristic then moves to the next demand on the demands list and repeats the above procedure. After completing all the demands list, the total power consumption resulting from routing all demands is calculated.

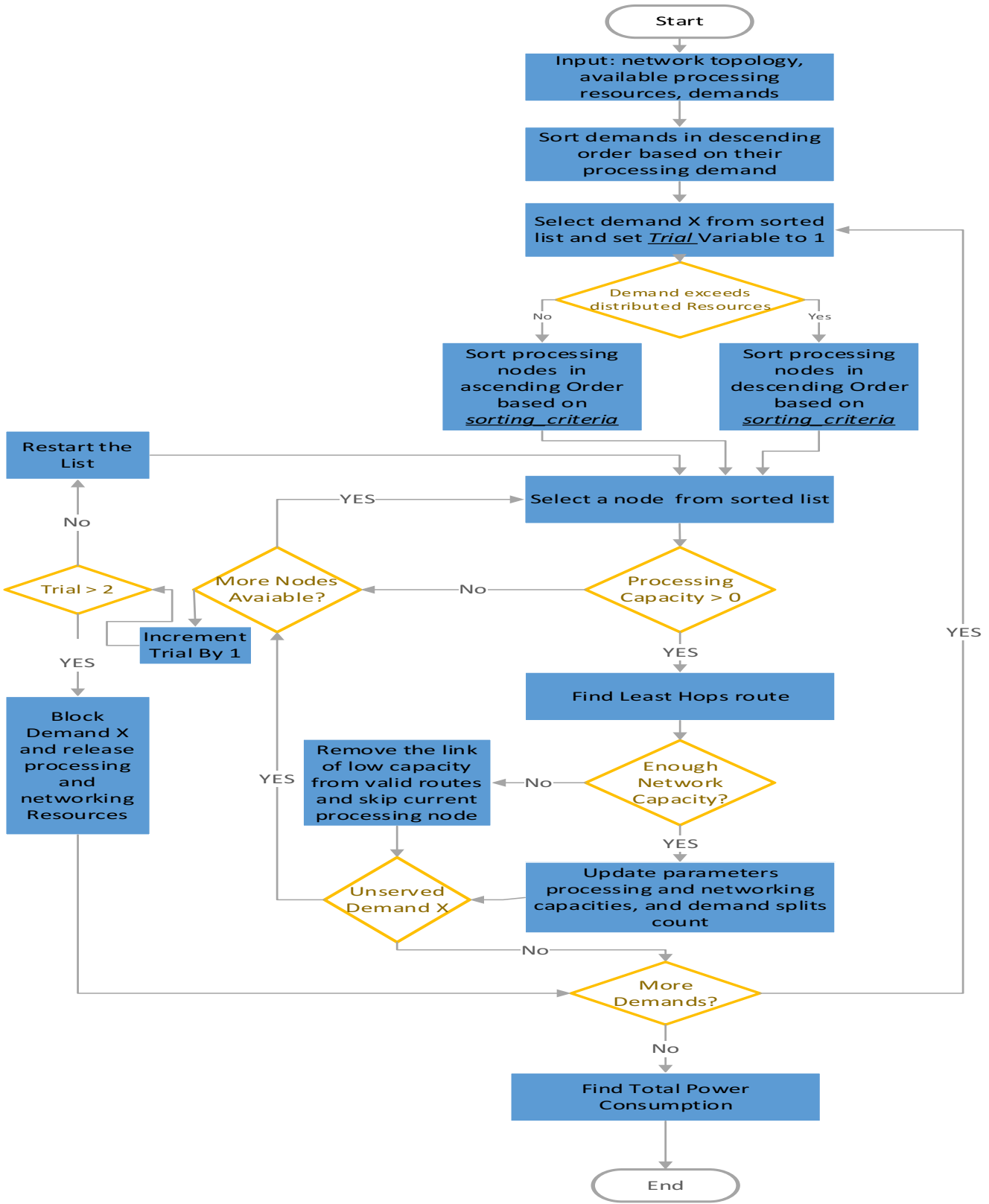

*Figure 24: Heuristic Flow Chart*

## A. Demand Size Variation

Figure 25 shows a comparison between the total power consumption of the heuristic and the MILP model results when considering a single request of varying size in the VEC scenario. Table 8 shows the gap in power consumption between the two. The heuristic approaches the model with a gap of 0%-15% for most of the demand sizes. This gap is a result of the heuristic sub-optimal selection of processing nodes, as seen in Figure 26, resulting from the sequential allocation of the processing based on the current status with no knowledge of the upcoming demands. For the heuristics there is a pattern of depleting the resources of the lower processing layers before allocating demands to higher processing layers. On the other hand, the MILP model allocates demands to edge nodes or a combination of vehicular and edge nodes even when vehicular nodes have more available resources.

*Table 8: MILP vs Heuristics Power Consumption Difference for Demand Size Variation*

| Traffic Demand (Mbps) | Diff (%) | Traffic Demand (Mbps) | Diff (%) |
|---|---|---|---|
| **2** | 0 | 14 | 11 |
| **4** | 23 | 16 | 3 |
| **6** | 16 | 18 | 11 |
| **8** | 13 | 20 | 3 |
| **10** | 14 | 30 | 0 |
| **12** | 0 | | |

The 30 Mbps demand is optimally processed in the cloud, both in the model and the Heuristics. In this case, the traffic demand exceeded the capacity available in the DSRC, which led to the list of candidates to be sorted in descending order, as stated in the flowchart. This arrangement pushed the cloud node to the front of the sorted list, and it was chosen as destination to serve the demand. If the processing nodes were sorted ascendingly, the cloud would have been at the end of the candidate list. As the heuristic would distribute the demand between the vehicles and edge node and when it tries to serve the remainder of the demand in the cloud, it finds the network capacity at the vehicles and edge node layers was already occupied by the traffic of distributed processing in the vehicular and edge layers.

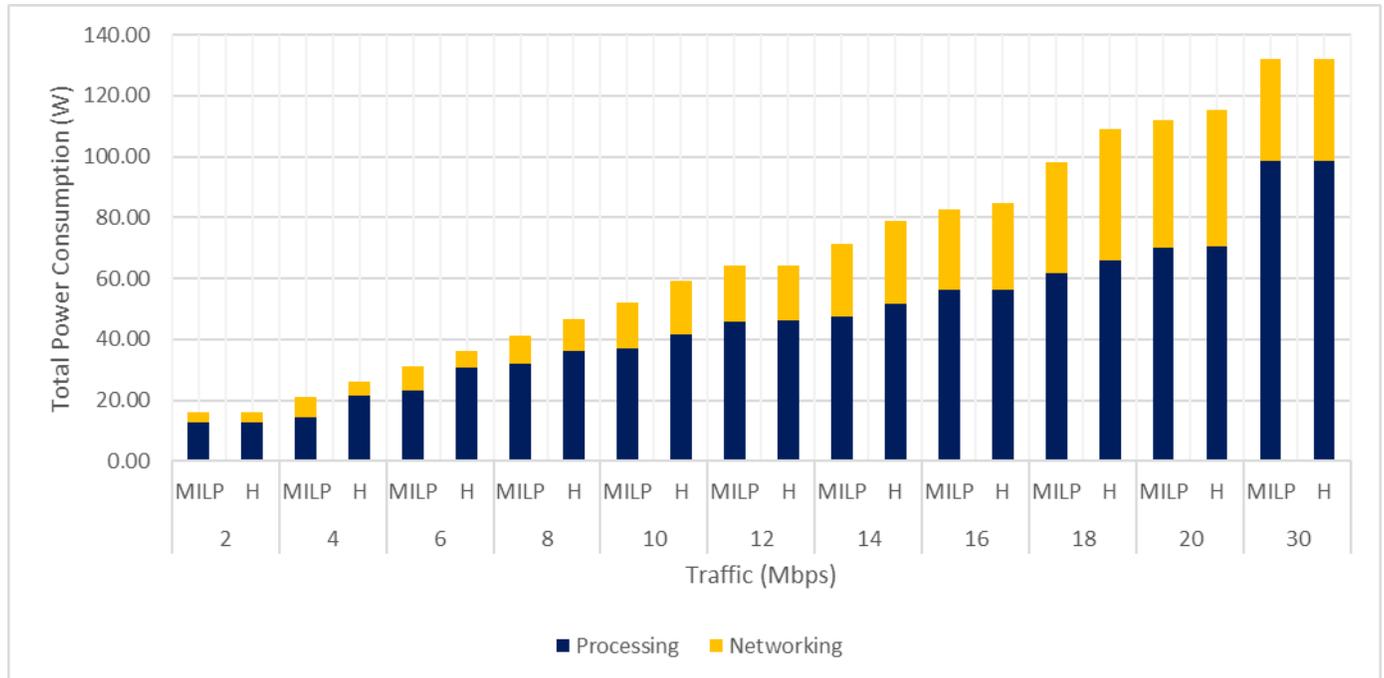

*Figure 25: Total power consumption when serving a single demand considering the VEC scenario (Heuristics and MILP)*

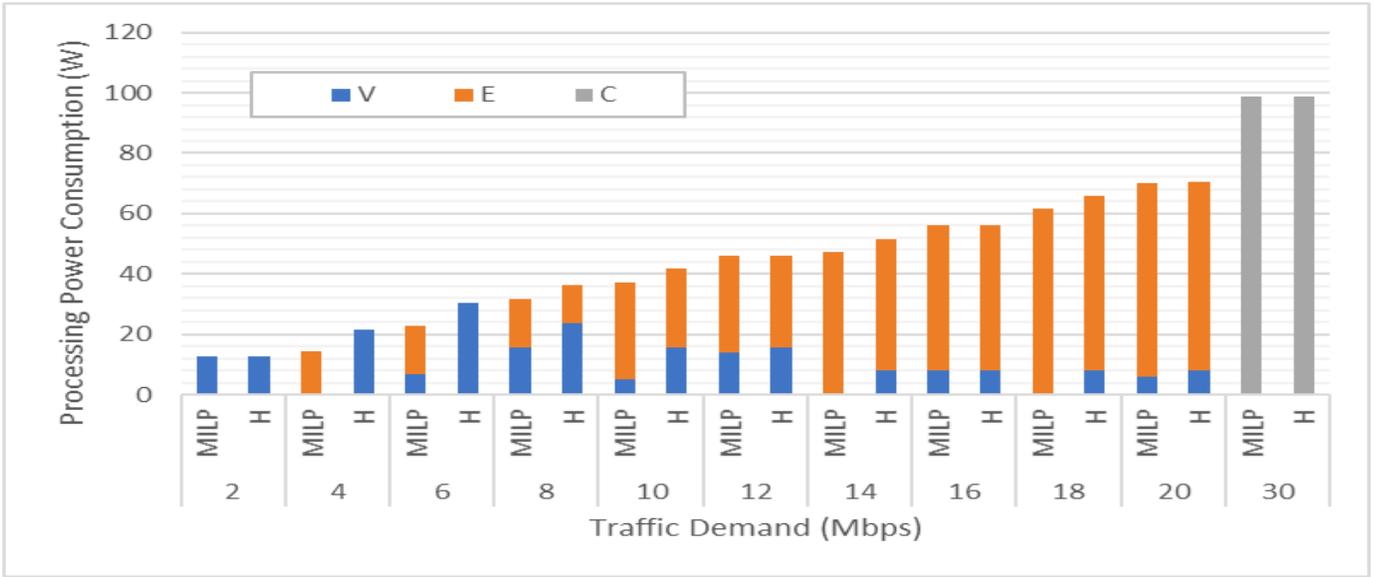

*Figure 26: Processing power consumption when serving a single demand considering the VEC scenario (Heuristics and MILP)*

## B. Processing Demand Splitting Limitations

As part of taking stock of the available resources, the heuristics counts the minimum number of nodes needed to process a specific demand in each processing layer. When it checks the candidate processing capacity, it also checks if the number of nodes required at the candidate layer is within the splitting limitation. If not, then it is seen as insufficient and the heuristic moves to the next candidate. Figure 27 shows a comparison with the results of the model in III.B, and Table 9 shows the gap between the power consumption values. For smaller demand values, the results are identical for all splits limitations. For demands of 3.5 Mbps and above, the results were identical for smaller number of splits, but as the number of splits increased, the heuristics produced higher power consumption. The reason for this is that as the limits on the number of processing nodes becomes larger, the possibility of serving in the vehicles increases. The heuristics only ensures that the number of processing nodes does not exceed the limit, while the MILP tends to consolidate the demand in a destination whenever possible, leading to fewer number of active nodes and lower power consumption.

*Table 9: Heuristic vs MILP Power Consumption Difference for Demand Splitting Limitation*

| Traffic Demand (Mbps)/Split | 1.5 | 2 | 2.5 | 3 | 3.5 | 4.5 | 5.5 |
|---|---|---|---|---|---|---|---|
| No Split | **0** | **0** | **0** | **0** | **0** | **0** | **0** |
| 2 | **0** | **0** | **0** | **0** | **0** | **0** | **23** |
| 3 | **0** | **0** | **0** | **0** | **23** | **21** | **23** |
| 4 | **0** | **0** | **0** | **0** | **23** | **21** | **16** |
| Unlimited | **0** | **0** | **0** | **0** | **23** | **21** | **16** |

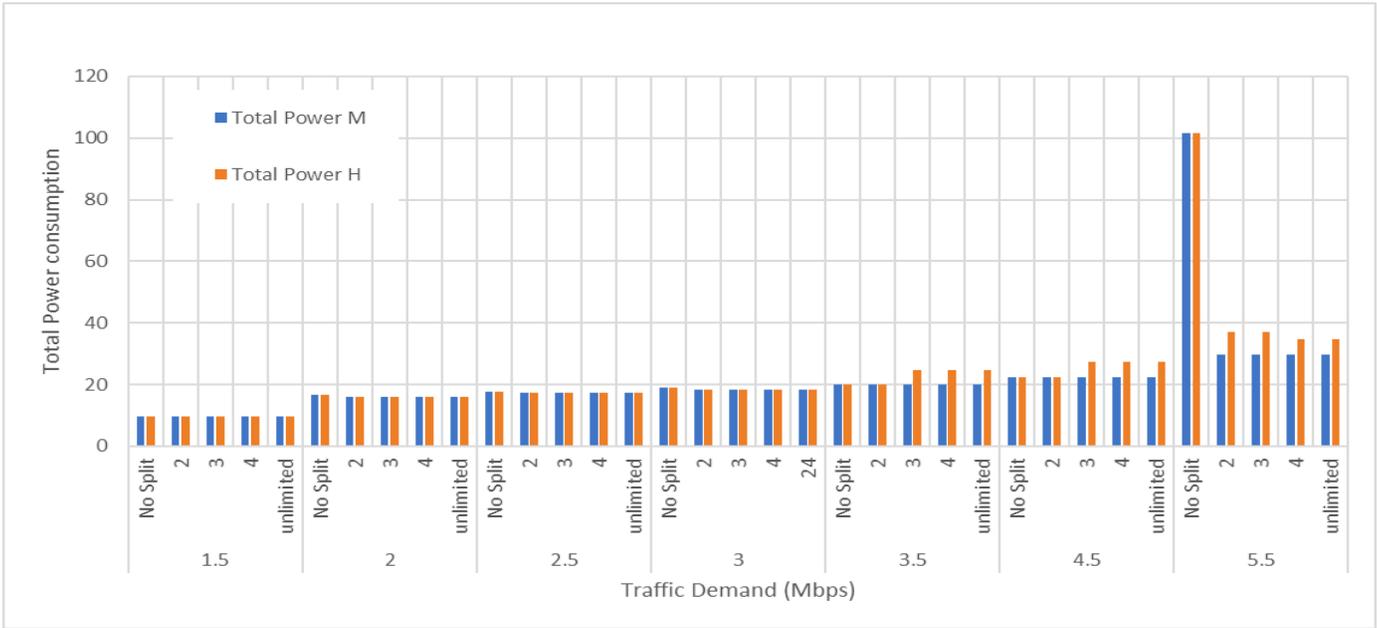

*Figure 27: Total power consumption with demand splitting limitations (Heuristics and MILP)*

## C. Proportional Traffic Assignment

*To implement the heuristics in this case, the same change that was made in the model in equation (16) was also made in the heuristics so that each destination receives traffic proportional to the processing demand it serves. Figure 28 shows the results of the heuristics for the proportional traffic in the V scenario, and it shows a complete match with the results of the MILP model. Figure 29 shows the results for the VE scenario and it shows higher power consumption in the heuristic, with the difference shown in*

Table 10. As was explained before, the difference in the power consumption comes from the sequential allocation of the demands in the heuristics.

Even though it is not shown, the VEC scenario had an exact match with the results of the V and VE scenarios in the heuristics, except for the 30 Mbps demand size, for which it made the same choice as the MILP and processed in the cloud.

*Table 10: Heuristic and MILP Power Consumption Difference for Proportional Traffic*

| Traffic Demand (Mbps) | Diff (%) | Traffic Demand (Mbps) | Diff (%) |
|---|---|---|---|
| **2** | 0 | 14 | 19 |
| **4** | 22 | 16 | 18 |
| **6** | 16 | 18 | 19 |
| **8** | 12 | 20 | 19 |
| **10** | 19 | 30 | 2 |
| **12** | 16 | | |

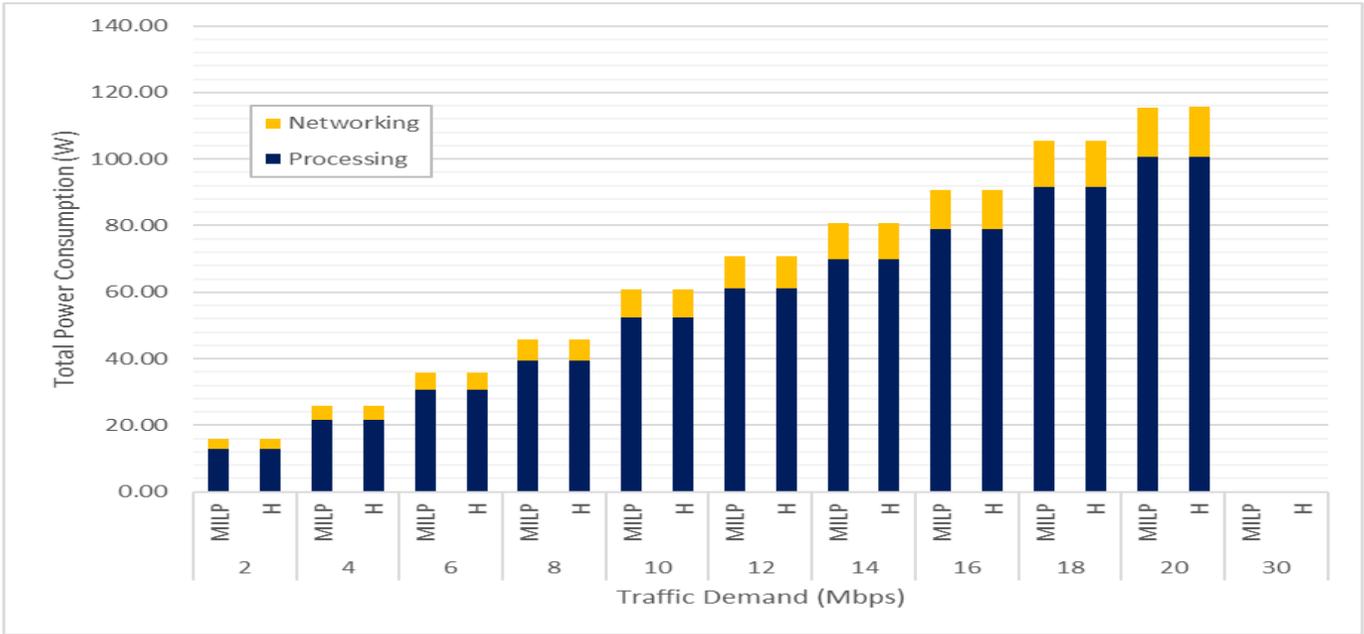

*Figure 28: Total power consumption of proportional traffic considering the V scenario (Heuristics and MILP)*

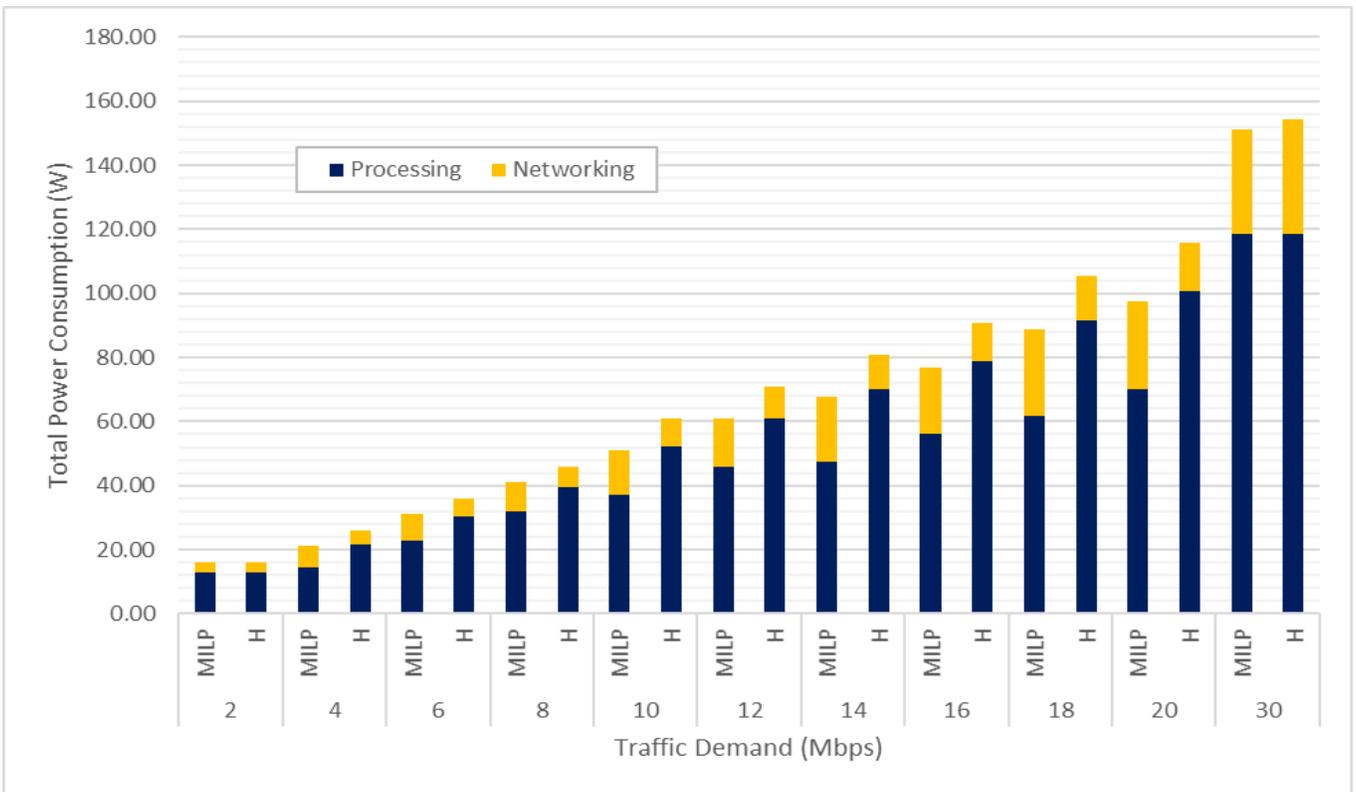

*Figure 29: Total power consumption of proportional traffic in VE scenario (Heuristics and MILP)*

### D. Multiple Demands Services

Figure 30- Figure 32 show a comparison of the power consumption for multiple demands as obtained from the model and the heuristics, all using VEC scenario. As before, the heuristics sequential approach gave priority for processing in the vehicles before moving to the edge and then the cloud. The distribution over more processing nodes and the choice of first sufficient route with least hops increases the power consumption in the heuristics. The behaviour is similar for low, medium, and high demand sizes. However, for the high demand sizes with the number of requests above 5, the demand requirements exceeded the capacities of the vehicles and edge layers. The heuristics sorted the candidates in descending order, therefore, the demands for these cases were served in the cloud, similar to the model. The higher power consumption of these cases is due mainly to the routing decisions and not processing. Table 11 shows the difference in the power consumption between the optimal solution (MILP) and the heuristic.

*Table 11: Heuristics vs MILP for Multiple Demands*

| Traffic Demand (Mbps) | Diff (%) |
|---|---|
| **Low** | 0-25 |
| **Medium** | 0-16 |
| **High** | 12-25 |

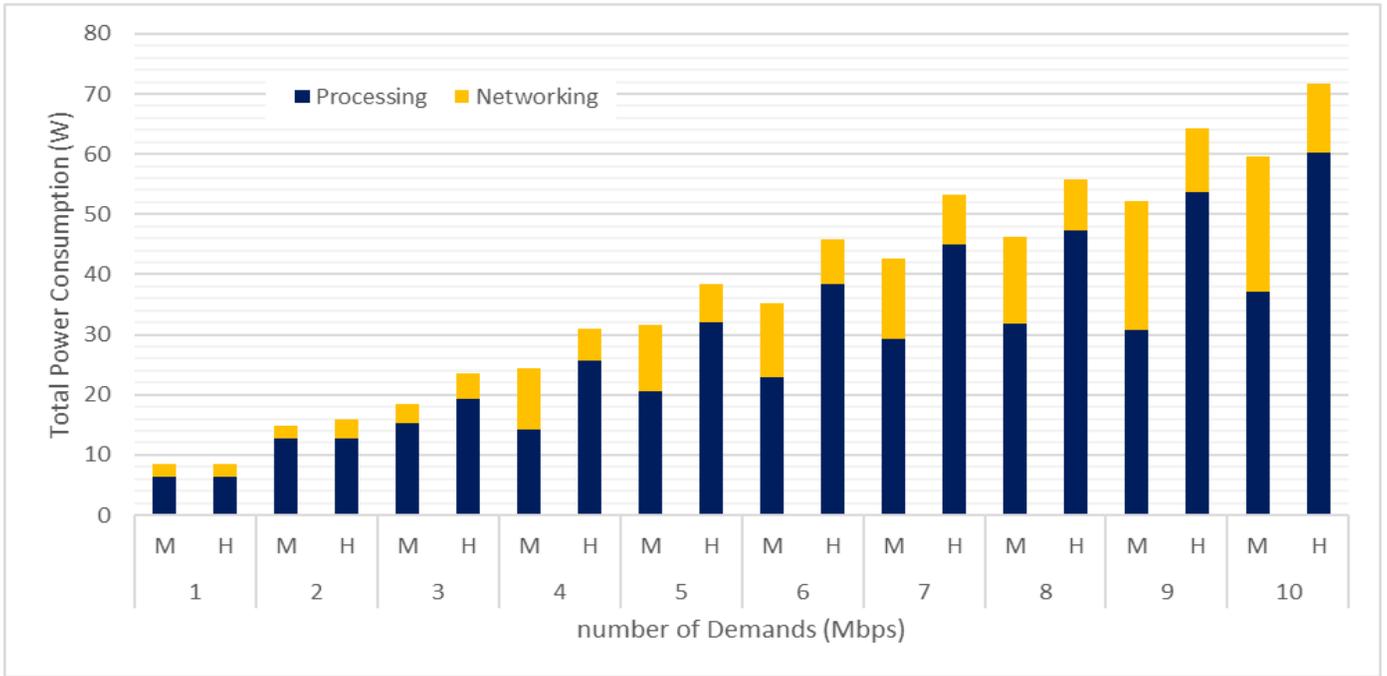

*Figure 30: Total power consumption for low requirements demands (Heuristics and MILP)*

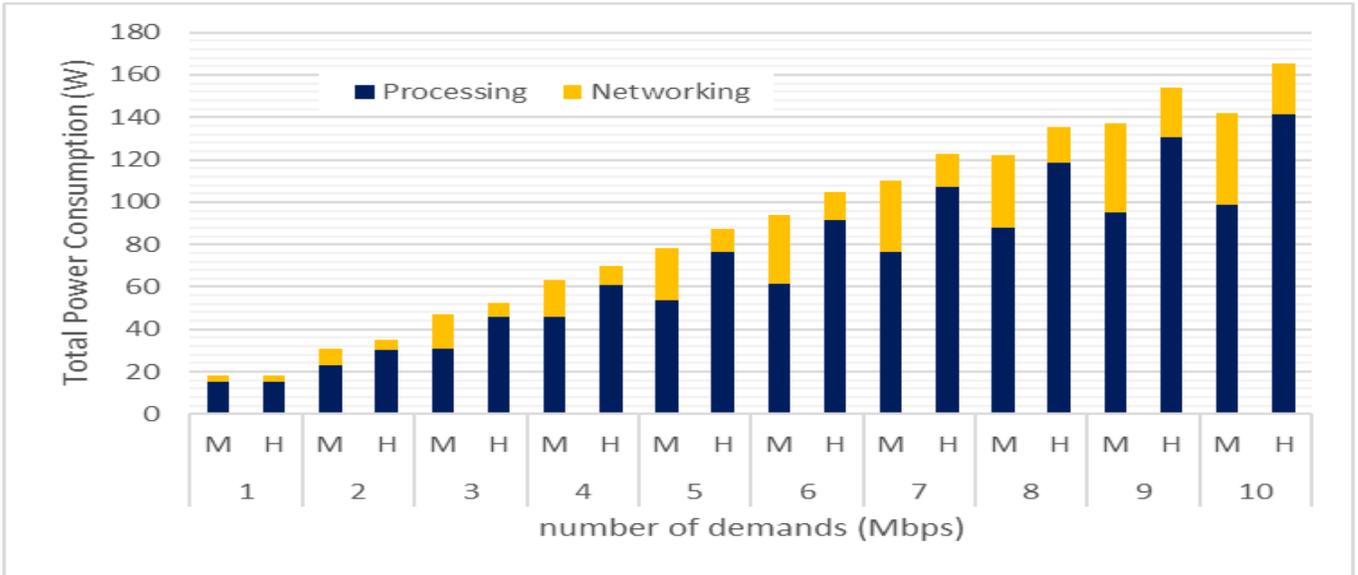

*Figure 31: Total power consumption for medium requirements demands (Heuristics and MILP)*

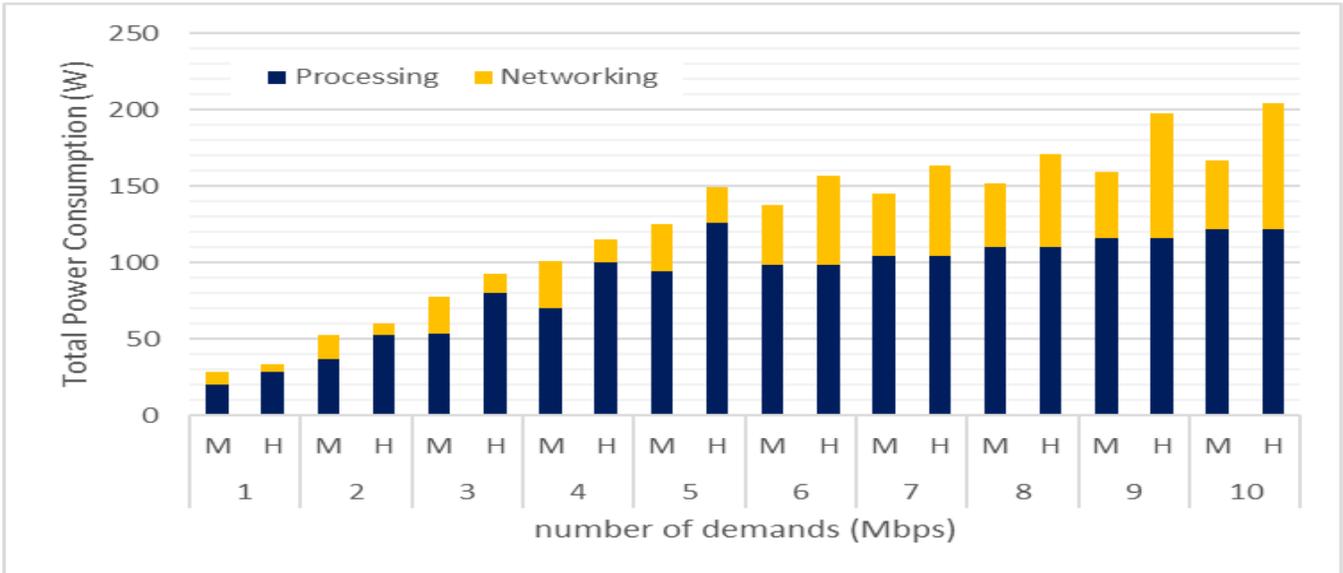

*Figure 32: Total power consumption for high requirements demands total power consumption (Heuristics and MILP)*

## V. Conclusions

This paper has investigated the use of underutilised computing resources in modern vehicles to create a processing layer, referred to as the vehicular cloud, in proximity of end users. The vehicular cloud complements conventional cloud computing and fixed edge computing in a distributed processing architecture. The architecture was modelled using a MILP model with the objective of minimizing the total power consumption. The results of the MILP model show that the energy efficiency of processing in vehicles compared to the cloud decreases as the size of the demand increases. Processing in a combination of vehicles and edge nodes results in average power savings of 6% compared to processing in the cloud for demands of traffic as high as 18 Mbps. The limited data rate of the vehicle wireless interfaces cannot support distributed processing in vehicles and edge nodes as the traffic is replicated to all processing destinations. Therefore, vehicular communication interfaces of higher data rate are essential to improve the utilisation of vehicular clouds. The results also illustrate that splitting a processing demand improves the energy efficiency of processing in the vehicles and edge nodes by 71%. Furthermore, the results show applications which require proportional traffic splitting among the processing destinations serving the demand. These applications can be more efficiently processed by vehicles and edge nodes, thus increasing the average power savings to 3%-16% compared to cloud processing, even for demands up to 20 Mbps. A real-time heuristic for allocating processing demands is developed based on insights from the model. The results show that the heuristic has comparable performance to the MILP model.

### Acknowledgment

This work was supported in part by the Engineering and Physical Sciences Research Council (EPSRC), in part by INTelligent Energy aware NETworks (INTERNET) under Grant EP/H040536/1, in part by SwiTching And tRansmission (STAR) under Grant EP/K016873/1, and in part by Terabit Bidirectional Multi-user Optical Wireless System (TOWS) project under Grant EP/S016570/1. All data is provided in the results section of this paper.

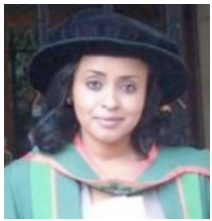

**Fatemah S. Behbehani** is currently a PhD student in the University of Leeds, UK in 2020 working on energy efficient vehicular cloud networks design. She has research interests in network architecture design, energy efficiency, network optimization, mixed integer linear programming and vehicular cloud and fog networks. She has published several papers in this area.

**Taisir E. H. EL-Gorashi** received the B.S. degree (first-class Hons.) in Electrical and Electronic Engineering from the University of Khartoum, Khartoum, Sudan, in 2004, the M.Sc. degree (with distinction) in Photonic and Communication Systems from the University of Wales, Swansea, UK, in 2005, and the PhD degree in Optical Networking from the University of Leeds, Leeds, UK, in 2010. She is currently a Lecturer in optical networks in the School of Electronic and Electrical Engineering, University of Leeds. Previously, she held a Postdoctoral Research post at the University of Leeds (2010– 2014), where she focused on the energy efficiency of optical networks investigating the use of renewable energy in core networks, green IP over WDM networks with datacenters, energy efficient physical topology design, energy efficiency of content distribution networks, distributed cloud computing, network virtualization and big data. In 2012, she was a BT Research Fellow, where she developed energy efficient hybrid wireless-optical broadband access networks and explored the dynamics of TV viewing behavior and program popularity. The energy efficiency techniques developed during her postdoctoral research contributed 3 out of the 8 carefully chosen core network energy efficiency improvement measures recommended by the GreenTouch consortium for every operator network worldwide. Her work led to several invited talks at GreenTouch, Bell Labs, Optical Network Design and Modelling conference, Optical Fiber Communications conference, International Conference on Computer Communications, EU Future Internet Assembly, IEEE Sustainable ICT Summit and IEEE 5G World Forum and collaboration with Nokia and Huawei.

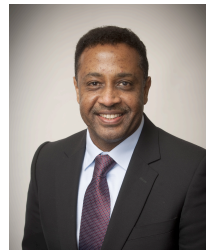

**Jaafar M. H. Elmirghani** is Fellow of IEEE, Fellow of the IET, Fellow of the Institute of Physics and is the Director of the Institute of Communication and Power Networks and Professor of Communication Networks and Systems within the School of Electronic and Electrical Engineering, University of Leeds, UK. He joined Leeds in 2007 having been full professor and chair in Optical Communications at the University of Wales Swansea 2000-2007. He received the BSc in Electrical Engineering, First Class Honours from the University of Khartoum in 1989 and was awarded all 4 prizes in the department for academic distinction. He received the PhD in the synchronization of optical systems and optical receiver design from the University of Huddersfield UK in 1994 and the DSc in Communication Systems and Networks from University of Leeds, UK, in 2012. He co-authored Photonic Switching Technology: Systems and Networks, (Wiley) and has published over 550 papers. He was Chairman of the IEEE UK and RI Communications Chapter and was Chairman of IEEE Comsoc Transmission Access and Optical Systems Committee and Chairman of IEEE Comsoc Signal Processing and Communication Electronics (SPCE) Committee. He was a member of IEEE ComSoc Technical Activities Council' (TAC), is an editor of IEEE Communications Magazine and is and has been on the technical program committee of 41 IEEE ICC/GLOBECOM conferences between 1995 and 2020 including 19 times as Symposium Chair. He was founding Chair of the Advanced Signal Processing for Communication Symposium which started at IEEE GLOBECOM'99 and has continued since at every ICC and GLOBECOM. Prof. Elmirghani was also founding Chair of the first IEEE ICC/GLOBECOM optical symposium at GLOBECOM'00, the Future Photonic Network Technologies, Architectures and Protocols Symposium. He chaired this Symposium, which continues to date. He was the founding chair of the first Green Track at ICC/GLOBECOM at GLOBECOM 2011, and is Chair of the IEEE Sustainable ICT Initiative, a pan IEEE Societies Initiative responsible for Green ICT activities across IEEE, 2012-present. He has given over 90 invited and keynote talks over the past 15 years. He received the IEEE Communications Society 2005 Hal Sobol award for exemplary service to meetings and conferences, the IEEE Communications Society 2005 Chapter Achievement award, the University of Wales Swansea inaugural 'Outstanding Research Achievement Award', 2006, the IEEE Communications Society Signal Processing and Communication Electronics outstanding service award, 2009, a best paper award at IEEE ICC'2013, the IEEE Comsoc Transmission Access and Optical Systems outstanding Service award 2015 in recognition of "Leadership and Contributions to the Area of Green Communications", the GreenTouch 1000x award in 2015 for "pioneering research contributions to the field of energy efficiency in telecommunications", the IET 2016 Premium Award for best paper in IET Optoelectronics, shared the 2016 Edison Award in the collective disruption category with a team of 6 from GreenTouch for their joint work on the GreenMeter, the IEEE Communications Society Transmission, Access and Optical Systems technical committee 2020 Outstanding Technical Achievement Award for outstanding contributions to the "energy efficiency of optical communication systems and networks". He was named among the top 2% of scientists in the world by citations in 2019 in Elsevier Scopus, Stanford University database which includes the top 2% of scientists in 22 scientific disciplines and 176 sub-domains. He was elected Fellow of IEEE for "Contributions to Energy-Efficient Communications," 2021. He is currently an Area Editor of IEEE Journal on Selected Areas in Communications series on Machine Learning for Communications, an editor of IEEE Journal of Lightwave Technology, IET Optoelectronics and Journal of Optical Communications, and was editor of IEEE Communications Surveys and Tutorials and IEEE Journal on Selected Areas in Communications series on Green Communications and Networking. He was Co-Chair of the GreenTouch Wired, Core and Access Networks Working Group, an adviser to the Commonwealth Scholarship Commission, member of the Royal Society International Joint Projects Panel and member of the Engineering and Physical Sciences Research Council (EPSRC) College. He has been awarded in excess of £30 million in grants to date from EPSRC, the EU and industry and has held prestigious fellowships funded by the Royal Society and by BT. He was an IEEE Comsoc Distinguished Lecturer 2013-2016. He was PI of the £6m EPSRC Intelligent Energy Aware Networks (INTERNET) Programme Grant, 2010-2016 and is currently PI of the EPSRC £6.6m Terabit Bidirectional Multi-user Optical Wireless System (TOWS) for 6G LiFi, 2019-2024. He leads a number of research projects and has research interests in communication networks, wireless and optical communication systems.